\documentclass[a4paper,11pt]{article}
\pdfoutput=1
\usepackage{jheppub}
\usepackage{graphicx,slashed,hyperref,xcolor,multirow}
\usepackage{amssymb}
\usepackage[normalem]{ulem}
\usepackage[utf8]{inputenc}
\usepackage{array}
\usepackage{ctable} 
\hypersetup{
   bookmarks=true,         
   unicode=true,          
   pdftoolbar=true,        
   pdfmenubar=true,        
   pdffitwindow=false,     
   pdfstartview={FitH},    
   pdftitle={My title},    
   pdfauthor={Author},     
   pdfsubject={Subject},   
   pdfcreator={Creator},   
   pdfproducer={Producer}, 
   pdfkeywords={keyword1} {key2} {key3}, 
   pdfnewwindow=true,      
   colorlinks=true,       
   linkcolor=blue,          
   citecolor=magenta,        
   filecolor=magenta,      
   urlcolor=cyan           
}

\def\lsim{\mathrel{\rlap{\lower4pt\hbox{\hskip1pt$\sim$}}
    \raise1pt\hbox{$<$}}}         
\def\gsim{\mathrel{\rlap{\lower4pt\hbox{\hskip1pt$\sim$}}
    \raise1pt\hbox{$>$}}}         

\newcommand{\gev}{\textrm{ GeV}}

\newcommand{\ttbar}{t\bar{t}}
\makeatletter

\newcommand{\fmslash}[2][0mu]{%
  \mathchoice
    {\fmsl@sh\displaystyle{#1}{#2}}%
    {\fmsl@sh\textstyle{#1}{#2}}%
    {\fmsl@sh\scriptstyle{#1}{#2}}%
    {\fmsl@sh\scriptscriptstyle{#1}{#2}}}
\newcommand{\fmsl@sh}[3]{%
  \m@th\ooalign{$\hfil#1\mkern#2/\hfil$\crcr$#1#3$}}
\makeatother

\newcommand{\beq}{\begin{equation}}
\newcommand{\eeq}{\end{equation}}
\newcommand{\bea}{\begin{eqnarray}}
\newcommand{\eea}{\end{eqnarray}}
\mathchardef\minus="002D

\usepackage{xcolor}

\newcommand{\sa}[1]{\textcolor{brown}{#1}}

\def\beq{\begin{equation}}
\def\eeq{\end{equation}}
\def\bea{\begin{eqnarray}}
\def\eea{\end{eqnarray}}

\title{Energy-peak based method to measure top quark mass via $B$-hadron decay lengths}

\author[a]{Kaustubh Agashe,}
\author[a]{Sagar Airen,}
\author[b]{Roberto Franceschini,}
\author[c]{Joseph Incandela,}
\author[d]{Doojin Kim,}
\author[a]{and Deepak Sathyan}

\affiliation[a]{Maryland Center for Fundamental Physics, Department of Physics, University of Maryland,
     College Park, MD 20742, USA}
\affiliation[b]{Universit\`{a} degli Studi and INFN Roma Tre, Via della Vasca Navale 84, I-00146, Rome}
\affiliation[c]{Department of Physics, University of California at Santa Barbara (UCSB), 5105 Broida Hall, Santa Barbara, CA 93106, USA}
\affiliation[d]{Mitchell Institute for Fundamental Physics and Astronomy,  Department of Physics and Astronomy, Texas A\&M University, College Station, TX 77843, USA}

\emailAdd{kagashe@umd.edu}
\emailAdd{sairen@umd.edu}
\emailAdd{roberto.franceschini@uniroma3.it}
\emailAdd{jincandela@ucsb.edu}
\emailAdd{doojin.kim@tamu.edu}
\emailAdd{dsathyan@umd.edu}

\preprint{
\begin{minipage}{5cm}
\begin{flushright}
UMD-PP-022-10 \\
MI-HEE-790
 \end{flushright}
\end{minipage}
}

\abstract{We develop a method for the determination of the top quark mass using the distribution of the decay length of the $B$-hadrons originating from its decay. This technique is based on our earlier observation regarding the location of the peak of the $b$ quark energy distribution. Such ``energy-peak" methods enjoy a greater degree of model-independence with respect to the kinematics of top quark production compared to earlier proposals. The CMS experiment has implemented the energy-peak method using associated $b$-jet energy as an approximation for $b$ quark energy. 
The new method uses $B$-hadron decay lengths, which are related to $b$ quark energies by convolution. The advantage of the new decay length method is that it can be applied in a way that evades jet-energy scale (JES) uncertainties. Indeed, CMS has measured the top quark mass using $B$-hadron decay lengths, but they did not incorporate the energy-peak method. Therefore, mismodeling of top quark transverse momentum remains a large uncertainty in their result. We demonstrate that, using energy-peak methods, this systematic uncertainty can become negligible. We show that with the current LHC data set, a sub-GeV statistical uncertainty on the top quark mass can be attained with this method. To achieve a comparable systematic uncertainty 
as is true for many methods based on exclusive or semi-inclusive observables using hadrons, we find that the quark-hadron transition needs to be described significantly better than is the case with current fragmentation functions and hadronization models.}

\begin{document}

\maketitle

\section{Motivation for a new method}

A precise measurement of the top quark mass is well-motivated for various reasons: see, for example, Section 1.2 of Ref.~\cite{Agashe:2013hma}.
Ideally, we should aim for a {method that is least sensitive to unknowns, e.g. the modeling of the kinematics of the production mechanism of top quarks at the LHC and other aspects of top quark physics which may affect the extraction of the top quark mass from the data}. {In this paper we refer to such a method as ``model-independent.''}

In this work we will assume that the top quark decays as in the SM into a bottom quark and a $W$ boson at leading order (neglecting CKM mixing):

\begin{eqnarray}
t & \rightarrow & b W\,
\end{eqnarray}

We are looking for a strategy that does not depend on knowledge of the fine details of the top quark's production mechanism.

Indeed, even if one assumes the SM to be the correct theory of Nature (in light of the absence of compelling evidence for new physics thus far!), there are still uncertainties in SM production itself, such as higher-order QCD effects or uncertainties in PDFs that have not yet been calculated. Therefore, it is very useful to have a mass measurement method that is model-independent. 

Furthermore, the goal of devising production-model-independent measurement methods of the top quark mass is motivated by the existence of models in which there is a potentially significant new physics contribution to a $t \bar{t}$ final state. 
Examples in the past included stealth stops in SUSY (i.e., decaying to top quark plus neutralino/gravitino and being almost degenerate with that combination: see, for example, Section 1.6.1.2 of Ref.~\cite{Agashe:2013hma} for more details) or those proposed to explain the ``anomaly" in the top quark forward-backward asymmetry seen at the Tevatron (see, for example, Section 1.4.3 of Ref.~\cite{Agashe:2013hma} for more details). Even though these particular proposals may be ruled out today, one cannot completely rule out the possibility of new physics in top quark samples.

However, most of the current methods for the measurement of the top quark mass assume the SM matrix element in production. For instance, this is true of {\em all} of the methods used in the latest Tevatron-LHC combination for top mass \cite{ATLAS:2014wva}, which reported a central value of 173 GeV with an uncertainty of $\sim 0.8$ GeV.\footnote{This feature is obvious for the ``matrix-element" method. On the other hand, any method that is based on a full reconstruction of the top quark decay, with either leptonic or hadronic $W$ boson decays, may appear to be model-independent but faces ambiguities that rely upon Monte Carlo to resolve. For example, combinatorial issues, such as determining which $b$-jet of two in $t \bar{t}$ events does one combine with a given reconstructed $W$ boson or in the case of leptonic $W$ boson decay, an additional discrete ambiguity in the longitudinal neutrino momentum. To address these issues, one relies on the simulation of these ambiguities, using the SM matrix element.} 
Other methods, currently too imprecise to be included in the Tevatron-LHC combination, have tried to maximize the degree of model independence that one can attain in mass measurement. Examples include methods based on kinematic endpoints~\cite{Chatrchyan:2013boa} and the location of the $b$-jet energy peak~\cite{CMS:2015jwa}, both of which have been implemented by the CMS collaboration, each with a central value of (approximately) 174 GeV with an 
uncertainty of $\sim 2$ GeV, which is consistent with the result from the combination.
To reduce the uncertainty, besides using more data than in these two early attempts, it is compulsory to add corrections for final state QCD radiation off the top quark to take into account off-shell top quark production. The inclusion of radiative corrections is by itself a source of model-dependence of these two methods, but at a quantitatively smaller (and better understood)  level of theory uncertainty compared to methods that are model-dependent from the very start. 
As things stand, it would then appear that the top quark mass is known model-{\em in}dependently with an uncertainty of 2 GeV, as opposed to 0.8~GeV (or less, see e.g. Ref.~\cite{CMS:2022kcl}) upon assuming pure SM production and combining TeVatron and LHC results, 
i.e., there seems to be room for improvement in the model-independent arena.

Irrespective of the above considerations of model-independent production, new methods for measurement of the top quark mass are valuable, especially if they are simpler and can be complementary to existing ones by involving different systematics.

Finally, once tested on the top quark, such new methods can be more confidently used to measure the masses of any new particles that may be discovered, either at the LHC or future colliders. 
This is especially the case for new particles decaying semi-invisibly, i.e., into visible/SM and invisible particles, where a full reconstruction of their decay chains on an event-by-event basis is not 
possible.

With the above motives in mind, we have developed a new method for top quark measurement 
that aims to be as model-independent as possible.
It is based on the general idea of an ``energy-peak'' as follows.  We consider the energy distribution of a massless particle in the laboratory frame arising from the two-body decay of a heavy particle produced unpolarized, whose boost distribution is arbitrary. Remarkably, the location of the peak in this energy distribution is identical to its single-valued energy in the rest frame of the parent, which depends on the parent's mass and that of the other decay product. 
It follows that if one knows the mass of the latter, the mass of the parent can be inferred from a measurement of the energy peak.
It's then clear that this energy-peak idea can furnish a measurement of the top quark mass via the energy of the bottom quark from its decay (given the precisely known value of the $W$ boson mass), which, based on the ``parent-boost-invariance'' alluded to above, is largely insensitive to details of the production mechanism of the top quark. As mentioned above, most other methods assume pure SM production which itself has uncertainties and certainly does not take into account the possibility of any Beyond Standard Model (BSM) contributions.

The original proposal along this line was to simply use the $b$-{\em jet} (i.e., inclusive) energy as a very good approximation to the bottom quark energy. This method has been successfully implemented by the CMS Collaboration~\cite{CMS:2015jwa}. However, this method is afflicted by the jet energy scale (JES) uncertainty. This drawback can be circumvented by using instead the decay length of a $B$-hadron (i.e., more exclusive) contained in the $b$-jet as a proxy for the bottom quark energy. In particular, one can make use of the fact that the $B$-hadron decay lengths are related to $b$ quark energies by a double convolution of the B hadron decay exponential and the b-quark fragmentation function.
On the flip side, the $B$-hadron decay length method is sensitive to hadronization and fragmentation dynamics and so complementary to the $b$-jet energy method is assured.
The CMS collaboration also performed a top quark mass measurement using $B$-hadron decay length~\cite{CMSlifetime}, where they assumed SM production. 

In this paper, our new proposal is to then appropriately dovetail these two methods to obtain a  ``best of both worlds'' determination of the top quark mass that is based on a measurement of the $B$-hadron decay length, but improved by the energy-peak concept as outlined above to obtain a result that is both free of JES uncertainty {\em and} largely independent of the top quark production model.

The remainder of the paper is structured as follows. In Section \ref{basic}, we review the invariant energy peak result and describe how it is used to measure the top quark mass using $b$-jet energy spectrum. Next, we describe how the same idea can be put to use to measure the top mass using $B$-hadron decay lengths and discuss the associated systematic uncertainties. In Section~\ref{practical}, we thoroughly discuss how to implement the new method to measure the top quark mass. In Section \ref{results}, we discuss the robustness of our method and present our detailed expectations for the statistical and various systematic uncertainties. Finally, in Section~\ref{conclude} we summarize our results and note possibilities for future improvements.

\section{Basic idea}
\label{basic}

We will now briefly summarize the invariant energy-peak idea and how it can be used to measure the top quark mass. First, we show how to obtain a model-independent measurement of the top quark mass by measuring the energy peak of the $b$ quark energy distribution in the laboratory frame. Then we describe how to measure the energy-peak of the $b$ quarks in two ways: (1) by using the  $b$-jet energy-peak as our proxy for the $b$ quark energy-peak and (2) by using the $B$-hadron decay length distribution as our proxy. We then discuss the associated systematic uncertainties and experimental effects of each method.

\subsection{$b$ energy-peak}

A new method to measure the top quark mass was proposed in Ref.~\cite{Agashe:2012bn} using only the directly observed laboratory-frame $b$-jet energy distribution.
The starting point is that in the rest-frame of the top quark, the bottom quark is ``monochromatic" in energy, with a value that is a simple function of the masses:
\bea
E_b^{\rm rest } & = & \frac{ m_t^2 - m_W^2 + m_b^2 }{ 2\;m_t } \,.
\label{Ebrest}
\eea

However, the top quark is boosted in the laboratory frame, and the magnitude of this boost and its direction with respect to that of the bottom quark in the top quark's rest-frame varies event-by-event. Thus, we obtain a distribution of the observed bottom quark energy. Now let's assume only that top quarks are produced (i) unpolarized, (ii) with a typical\footnote{Further details on the conditions of the boost distribution are given in Ref.~\cite{Agashe:2012bn}.}
boost distribution, and (iii) they undergo the above two-body decay without final state radiation (FSR). In particular, there is no hard gluon emission off the bottom quark. In this case, it was shown~\cite{Agashe:2012bn} that the location of the peak in the bottom quark energy distribution is given by
\begin{eqnarray}
E_b^{\rm lab,\:peak} & = & \frac{ m_t^2 - m_W^2 + m_b^2 }{ 2\;m_t }.
\label{Emode}
\end{eqnarray}
This value is identical to the bottom quark energy in the rest-frame of the top quark: see Eq.~(\ref{Ebrest}).\footnote{This is just a specific case of a general result for two-body decays which was shown in Ref.~\cite{Agashe:2012bn}. It was then applied to the measurement of masses of 
hypothetical new particles in Ref.~\cite{Agashe:2013eba}.} Thus, assuming $m_W$ from an independent measurement, we get a simple measurement of the top quark mass from the bottom quark ``energy-peak". The energy-peak method is quasi-model-independent, i.e., it does not depend on many details of the production mechanism of the top quarks, such as the presence of 
initial state radiation (ISR), whether the top quark is produced singly or in pairs via SM production or via a new physics effect. It is also not affected by uncertainties in parton distribution functions (PDFs).

Of course, the bottom quark hadronizes, resulting in a $b$-jet, which is what is experimentally detected. However, being a largely 
inclusive quantity, $b$-jet energy is a good approximation to the original bottom quark energy, i.e., heuristically speaking, the probability for hadronization is unity.

\subsection{$B$-hadron decay length} \label{B-decay-length}
We expect that the features in the energy distribution of the bottom quark (such as the peak mentioned above) can be ``translated" into features in the decay lifetime/length of the $B$-hadron contained inside the $b$-jet. Let us now explore this carefully.
For simplicity, let's assume here that the bottom quark can only decay to one kind of $B$-hadron. (This is for illustration only and is not done in the actual analysis.) First, take the case of fixed energy of bottom quark, denoted simply by $E_b$, i.e., dropping the superscript ``lab" henceforth in this section. Even in this case, it is clear that we obtain a distribution of $B$-hadron energies (denoted by $E_B$) originating from it. Namely, the probability for the $B$-hadron energy to be between the values $E_B$ and $E_B + dE_B$ is given
 by the fragmentation function, denoted here by $D \left( E_B; E_b \right)$, where we are keeping track of a possible $E_b$ dependence. We normalize $D$ such that 

\bea
\int d E_B \; D\left(E_B; E_b\right) = 1\;
\hbox{for any (fixed)} \; E_b\,.
\eea
Thus, as shorthand, denoting the probability distribution functions or pdf's for the energies $E_b$ and $E_B$ by $f$ and $F$, respectively, we have the   ``convolution" relation:
\bea
F \left( E_B \right) & = & \int d E_b \; f \left( E_b \right) D \left( E_B ; E_b \right)\,,
\label{pdf_EB}
\eea
where $\int d E_b \; f \left( E_b \right) = 1$ (as is true for $F$ and other pdf's discussed below).
Note that from the discussion above we have:
\bea
\hbox{maximum of} \; f \left( E_b \right) & = & f \left( E_b^{ \rm rest } \right), \; {\rm with}~ E_b^{ \rm rest}~ \hbox{given by Eq.~(\ref{Ebrest})}\,
\label{pdf_Eb_peak}
\eea
While for fixed $E_B$, given the probabilistic nature of the decay process, we will
obtain a distribution of
decay 
times of the bottom quark, whose mean value is given by:
\begin{eqnarray}
\tau_B^{\rm lab} & = & \gamma_B {\tau}_B^{\rm rest} \nonumber \\
 & = & \frac{ E_B }{ m_B } \tau_B^{\rm rest} \,,
\end{eqnarray}
where $\tau_B^{\rm rest}$ is the proper lifetime of the $B$-hadron.\footnote{$\tau_B^{\rm rest}$ is in the $B$-hadron's rest frame, not
that of the top quark! Just for clarity, we re-introduce the superscript ``lab" on the LHS here.}
It follows that in the laboratory frame (for a fixed $B$-hadron energy) the mean decay length for a collection of $b$ quarks with fixed energy is given by:
\begin{eqnarray}
\lambda_B & = v_B \tau^{\rm lab}_B = & \beta_B c\cdot \gamma_B  \tau^{\rm rest}_B \nonumber \\
& = &  \sqrt{ 1 - \left( \frac{ m_B }{  E_B } \right)^2 }c \cdot \frac{ E_B }{ m_B } \tau^{\rm rest}_B \,
\label{lambda}
\end{eqnarray}

For relativistic $b$ quarks, 
$E_{B} \gg m_{B}$, and the above equation becomes 
\begin{eqnarray}
\lambda_B 
%
%
& \approx & c \frac{ E_B }{ m_B } \tau^{\rm rest}_B \left[  1 + \mathcal{O} \left( \left( \frac{ m_B }{  E_B } \right)^2 \right)\right] \,
\label{lambdarelativistic}
\end{eqnarray}
In other words, the probability for the $B$-hadron to have a decay length  
between the values of $L_B$ and $L_B + d L_B$ is 
proportional to $\exp \left( - L_B / \lambda_B \right) d L_B$.
Since $E_B$ varies event-by-event, as per Eq.~(\ref{pdf_EB}), it's clear that $\lambda_B$ itself has probability 
given by $g \left( \lambda_B \right) d \lambda_B ~ = ~ F \left( E_B \right)  d E_B$. Thus:
\bea
g \left( \lambda_B \right) & = &   \frac{ F \left( E_B \right) } { \frac{ d \lambda_B }{ d E_B } } \label{g_F_exact} 
\\
& \approx & F \left( E_B \right) \frac{ m_B }{ c \tau^{ \rm rest } _B } \,
\label{g_F_approx}
\eea
where in the second line, we have made the relativistic approximation in Eq.~(\ref{lambdarelativistic}) to simplify the equations in the discussion below. We will retain the full form of Eq.~(\ref{lambda}) in the calculations presented in later sections. 
We will have another convolution (via the decay exponential) to get the $L_B$-distribution from that of $\lambda_B^{ \rm lab }$.

Combining all of the above ingredients, we finally have an (albeit somewhat schematic) expression for the probability distribution $G \left( L_B \right)$ for the quantity that we actually measure in this method:
\bea
G \left( L_B \right) & = & \int d \lambda_B  \Big[ \frac{ g \left( \lambda_B
\right) }{ \lambda_B }  \Big]
\exp \left( - \frac{ L_B }{ \lambda_B
}
\right) \\ 
& \approx & \int d E_B \frac{ F \left( E_B \right) }{ E_B } \frac{ m_B }{ c \tau_B^{ \rm rest } } 
\exp \left( - \frac{ L_B m_B }{ c \tau_B^{ \rm rest } E_B } \right) \\
& = & \int d E_B \int d E_b \; f \left( E_b \right) D \left( E_B; E_b \right) 
\frac{ m_B }{ c \tau_B^{ \rm rest } E_B }
\exp \left( - \frac{ L_B m_B }{ c \tau_B^{ \rm rest } E_B } \right) 
\label{double-convolution}
\eea
where in the first line the factor of $1 / \lambda_B$ is from proper normalization of the decay exponential; 
the second line invokes the relativistic approximation and in the third line, we have related the distribution of $E_B$ to that of $E_b$ as per Eq.~(\ref{pdf_EB}).
For the sake of clarity, we summarize relevant variables and pdf's here:
\bea
G \left( L_B \right) & \rightarrow & \hbox{pdf of decay length of $B$-hadron}, \; L_B \nonumber \\ 
f \left( E_b \right ) & \rightarrow & \hbox{pdf of energy of bottom quark}, \; E_b \nonumber \\
\hbox{location of peak of} \; f \left( E_b \right) & \rightarrow & \frac{ m_t^2 - M_W^2 + m_b^2 }{ 2 \; m_t }
 \nonumber \\
D \left( E_B; E_b \right) & \rightarrow & \hbox{bottom quark fragmentation function} \nonumber \\
\tau_B^{ \rm rest } & \rightarrow & \hbox{mean decay lifetime of $B$-hadron in its rest frame} \nonumber \,.\\
\eea
Thus, we can hope to deduce the $f \left( E_b \right)$ distribution and hence the location of its peak, by twice ``de-convolving" the observed decay length: once to remove the exponential decay law and a second time to remove the fragmentation.
Equivalently, the simple and robust predictions for properties of the distribution of $f \left( E_b \right)$ (for example, the location of the peak in terms of masses and a few other features) can then ``materialize" in the distribution of $L_B$ via Eq.~(\ref{double-convolution}).
A concrete strategy for achieving this is discussed in Section \ref{practicaldecay}.

The CDF~\cite{Abulencia:2006rz} and CMS~\cite{CMSlifetime} Collaborations have determined the top quark mass based on 
measurements of the decay length using one of the original proposals in 
Ref.~\cite{Hill:2005zy}.
In particular, the measured decay length in the transverse plane, $L^{\rm mean}_{ B, \; x y }$, is used and they relate the top quark mass to the {\em mean} of the distribution of this quantity.
It is not possible to make a simple and robust (i.e., model-independent) statement 
about the distribution of $L^{\rm}_{ B, \; x y }$ as was just presented for $L_B$ above.\footnote{Note that the energy-peak idea in Ref.~\cite{Agashe:2012bn} was not known when these analyses were performed!}
In these analyses it is assumed that the top quark is produced via the SM matrix element and simulations were then performed to obtain the $L^{\rm mean }_{ B, \; x y }$ as a fully numerical ``function" of the top quark mass. Because of this, the analyses are
maximally exposed to uncertainties in the modeling of all aspects of top quark production and decay, and in the transition from quarks to hadrons. On the contrary, with the energy-peak method, we expect to track the top quark mass down to the peak of the $\lambda_{B}$ distribution, which can be predicted without simulation. The exact relation between this peak and the top quark mass does however still require complete knowledge of quark-hadron transition dynamics. The uncertainties in this knowledge will result in uncertainties on the top mass that are in common with the $L^{\rm mean }_{ B, \; x y }$ method and will be discussed in detail below.

\subsection{Systematics and experimental effects}
The two methods described in the previous two subsections have very different systematics. The $b$-jet energy-peak method was applied by the CMS Collaboration to measure the top quark mass \cite{CMS:2015jwa} where it was found that the jet energy scale was a major source of uncertainty. There are of course many other experimental effects resulting from the selection criteria, background estimation, etc., which were found to be under control.

The new method using $B$-hadron decay lengths involves some very different experimental effects and systematics. As mentioned earlier $B$-hadron decay lengths have been used by CDF and CMS to measure the top quark mass. Because our proposal makes use of decay lengths it should be sensitive to the same experimental details like secondary vertex reconstruction, tracker resolution, and so on. These systematics were under control in the previous implementation using decay lengths, and we expect the same for our new method. The biggest sources of uncertainty in the previous work ~\cite{Abulencia:2006rz, CMSlifetime} were due to the sensitivity to the hadronization model and top quark production modeling. As discussed in the previous section, the new method we propose should have a negligible top quark production modeling uncertainty while the uncertainty associated with hadronization remains. Clearly, the new proposal described in Section~\ref{B-decay-length} is not plagued by JES uncertainty. Therefore, the two methods inspired by the invariant energy-peak result complement each other well because of the different systematics. 
 
Now, we will elaborate further on how the model-independent nature of the two methods shields them from the uncertainties associated with production modeling. We will consider the following three small and independent, dimensionless parameters.
First, $\delta_{\rm prod}$ will denote the relative uncertainty in our knowledge of the unpolarized part of the cross-section for top quark production. This includes effects such as limited precision in the calculation of QCD production (due to PDF uncertainties or absence of corrections at higher-order in $\alpha_s$ etc.) as well as the potential contribution of unknown BSM physics. Obviously, focusing on the two-body decay of the top quark, one can see that  $\delta_{\rm prod}$ will not change the energy peak of Eq.~(\ref{Emode}).

Next, suppose that $f_{\rm pol}$ is the fraction of the total production cross-section that corresponds to polarized top quarks. If the top quark is polarized, then, the proof for energy-peak of Eq.~(\ref{Emode}) does not strictly hold. In particular, the peak position can be $O(1)$ different if the top quarks are, say, purely left or right-handed. Combining the polarized and unpolarized parts of the production can therefore result in a shift of the energy peak from Eq.~(\ref{Emode}) by $O \left( f_{\rm pol} \right)$. Of course, the new position can be calculated given the matrix element in production, i.e., the boost distribution of top quarks and degree of top quark polarization.

We will focus on pair-produced top quarks so that in the SM, $f_{\rm pol}$ originates from SM $Z$ boson exchange and is thus negligible, namely, $\sim \left( \alpha_{\rm EW}^2 / \alpha_s^2 \right) \times \hbox{the ratio of gluon to} \; q \bar{q}$ PDFs. Again, the shift in the energy peak resulting from this $Z$ exchange is in principle known. Thus, any relevant $f_{\rm pol }$ actually comes from new physics and hence is (a priori) unknown and constitutes an error to the prediction of the energy-peak location.

Finally, consider the contribution from hard gluon radiation off a bottom quark, which is part of an effective {three}-body decay of the top quark.\footnote{Clearly, the energy associated with soft radiation off the bottom quark is included in the $b$-jet energy so that this case is still effectively a two-body decay.} For such a decay, even in the rest frame of the top, we get a non-trivial, ``delocalized" (i.e., not the $\delta$-function of a strictly two-body decay) energy spectrum  for the bottom quark. This distribution can be calculated assuming the SM matrix element {\it in the top quark decay}, which is, of course, independent of the top quark production mechanism. Once again the proof of the energy-peak proxy for a single-valued $b$-jet energy in the top quark rest frame does not apply and the new peak position will depend on the top quark boost distribution which can be $O(1)$ different than in Eq.~(\ref{Emode}). Of course, if we know the matrix element in production, then the location of the new energy peak is calculable. Let us denote by $\epsilon_{\rm FSR}$ the size of this FSR contribution relative to the two-body decay of the top quark and denote by $\Delta_{\rm QCD}$ the shift in the energy peak caused by FSR. The contribution $\epsilon_{\rm FSR}$ is  suppressed by (perturbative) $\sim \alpha_s / \pi$ ($\sim 5 \%$) and perhaps even further if we impose a jet-veto (for the extra radiation). Upon adding this to the leading-order (two-body), the energy peak might thus shift from Eq.~(\ref{Emode}) by $O \left( \epsilon_{\rm FSR} \right)$.

Combining all the above three effects, we get the following schematic result for the observed lab-frame energy peak
\begin{eqnarray}
E_b^{\rm lab, mode} & = & \frac{ m_t^2 - m_W^2 + m_b^2 }{ 2 \; m_t } \left\{ 1 +  \epsilon_{\rm FSR} 
\Big[  \Delta_{\rm QCD} + O \left( \delta_{\rm prod} \right) \Big] + O \left( f_{\rm pol}  \right) \right\}\,,
\label{Emodeerror}
\end{eqnarray}
where we assume that the schematic terms inside the curly brackets have unknown O(1) coefficients.\footnote{The exact coefficients are computed in Ref.~\cite{Agashe:2016bok}.} Because $\Delta_{\rm QCD}$ can be calculated precisely, it does not affect the uncertainty in $m_t$.
Thus, the relative uncertainty in the top quark mass using the energy-peak method is: 
\begin{eqnarray}
{\left.\frac{\delta m_t}{m_t} \right|_{\rm energy-peak}} & = & 
\Big[ \mathcal{O} \left(  \epsilon_{\rm FSR} \delta_{\rm prod} \right) + \mathcal{O} \left( f_{\rm pol}   \right) \Big].
\label{Ebpeak error}
\end{eqnarray}
On the other hand, the relative error in most of the other methods, where it is crucial to  assume the SM matrix element in production to produce templates for the observables of interest can be written as:
\begin{eqnarray}
{\left.\frac{\delta m_t}{m_t} \right|_{\rm SM-templates}} & = & \mathcal{O} \left( \delta_{\rm prod} \right)\,.
\end{eqnarray}
Therefore, assuming top quarks are produced unpolarized, {i.e., $\mathcal{O}(f_{\rm pol})\approx 0$}, both the $b$-jet energy-peak method and the $B$-hadron decay length method via an energy-peak approach feature a rough ``double"-suppression as far as uncertainties in the production mechanism are concerned whereas there is only a single level of suppression for matrix element-based methods.

\section{General description of the analysis}
\label{practical}

\subsection{Signal and background definition}

We take our baseline selection criteria for the $t \bar{t}$ sample along the lines of what has already been used by the CMS and ATLAS Collaborations~\cite{Chatrchyan:2012ea,ATLAScuts}; namely 
{\em opposite}-flavor dilepton final states, \begin{equation}
pp\to e^+ \; \mu^- \; \textrm{ or }\; e^- \mu^+ + {\rm jets},  \label{eq:ttbar2emu}\end{equation}
where the observed collection of jets contains two $b$-tagged jets.
This channel has the best chances to reduce possible issues with our method, as some of the requirements for the separation of $t\bar{t}$ events from the background are not necessary. For instance, neither a $Z$-mass veto for the invariant mass of dileptons (which is needed for same-flavor dileptons) nor a missing transverse energy (denoted by MET or $E_T^{\rm miss}$) cut are required.
The absence of these requirements makes the selection more inclusive than that in other channels, and hence the chance to preserve the energy peak in Eq.~(\ref{Emode}) is maximized. Despite such a light set of event selection criteria, the choice of the opposite-flavor leptonic final state gives a high level of signal purity. Indeed, the resulting $S/B$ observed at the LHC for  $\sqrt{s}=$~7 TeV   
is $\sim 20$, as seen in Table 1 of Ref.~\cite{Chatrchyan:2012ea}.

In addition to the ultra-clean opposite-flavor, fully-leptonic channel, we also consider {\it same}-flavor fully leptonic and general semi-leptonic final states 
\begin{eqnarray}
pp &\to& e^+ \; e^- \; \textrm{ or }\; \mu^- \mu^+  + {\rm jets}\,, \\
pp  &\to& \ell^\pm + {\rm jets}\,,
\label{eq:ttbar2eemumu:semileptonic}\end{eqnarray}
where $\ell=e,\mu$. These channels need additional event selection criteria to reach a good level of $S/B$~\cite{CMS:2022kcl}. These channels bring additional events that increase overall statistics, and so a balance between purity and efficiency needs to be taken into consideration to achieve an optimal result. We address this aspect of the analysis below.

In any case,  some background contamination is expected and needs to be dealt with to achieve the best precision in the top mass extraction. Obviously, the $b$-jet energy distribution from background processes is not correlated with the top quark mass. One strategy to deal with $b$-jets from the background is to subtract the expected shape using a Monte Carlo template. We will work under the assumption that this subtraction can be successfully administered based upon the limited effect of residual background uncertainty achieved in recent analyses, e.g., Ref.~\cite{CMS:2022kcl}.

Some remarks are in order about the  backgrounds to the processes we are focusing on here. The dominant background turns out to be from {\em single} top quark production processes,
$$d\bar{u} \to W^* \to b t, \; qb\to q' t\;,\;
\textrm{and}\; gb \to W t,$$
all of which contain an on-shell top quark (or anti-quark).
Thus, one might naively expect the energy peak from this background to be the 
same as our signal in Eq.~\eqref{eq:ttbar2emu} from pair production. This turns out not to be the case because some of the requirements for the production mechanism model-independence are not satisfied by single top production.
First of all, the energy of the $b$-jet (out of a total of up to two in the event) coming from top decay {\em is} of course correlated with the top quark mass, but since the production of a single top quark always involves a $W$ boson, the top quark is polarized. 
Thus, we expect a shift in the energy peak with respect to Eq.~(\ref{Emode}). The shift is calculable and will be fully removed when the background templates match their actual shape in the data. Still, as such removal may be imperfect, it is important to keep in mind any possible adverse effects of having $b$-jets that do not originate from QCD pair production of top quarks.
Secondly, and maybe more importantly, the properties of the other $b$ jet emerging in single top production are not at all related to the top quark mass. Therefore, the inclusion of such $b$-jets in the analysis would result in a significant bias in the extracted top quark mass beyond the inherent sources listed in Eq.~(\ref{Emodeerror}).

\begin{figure}
    \centering
    \includegraphics[width=0.59\textwidth]{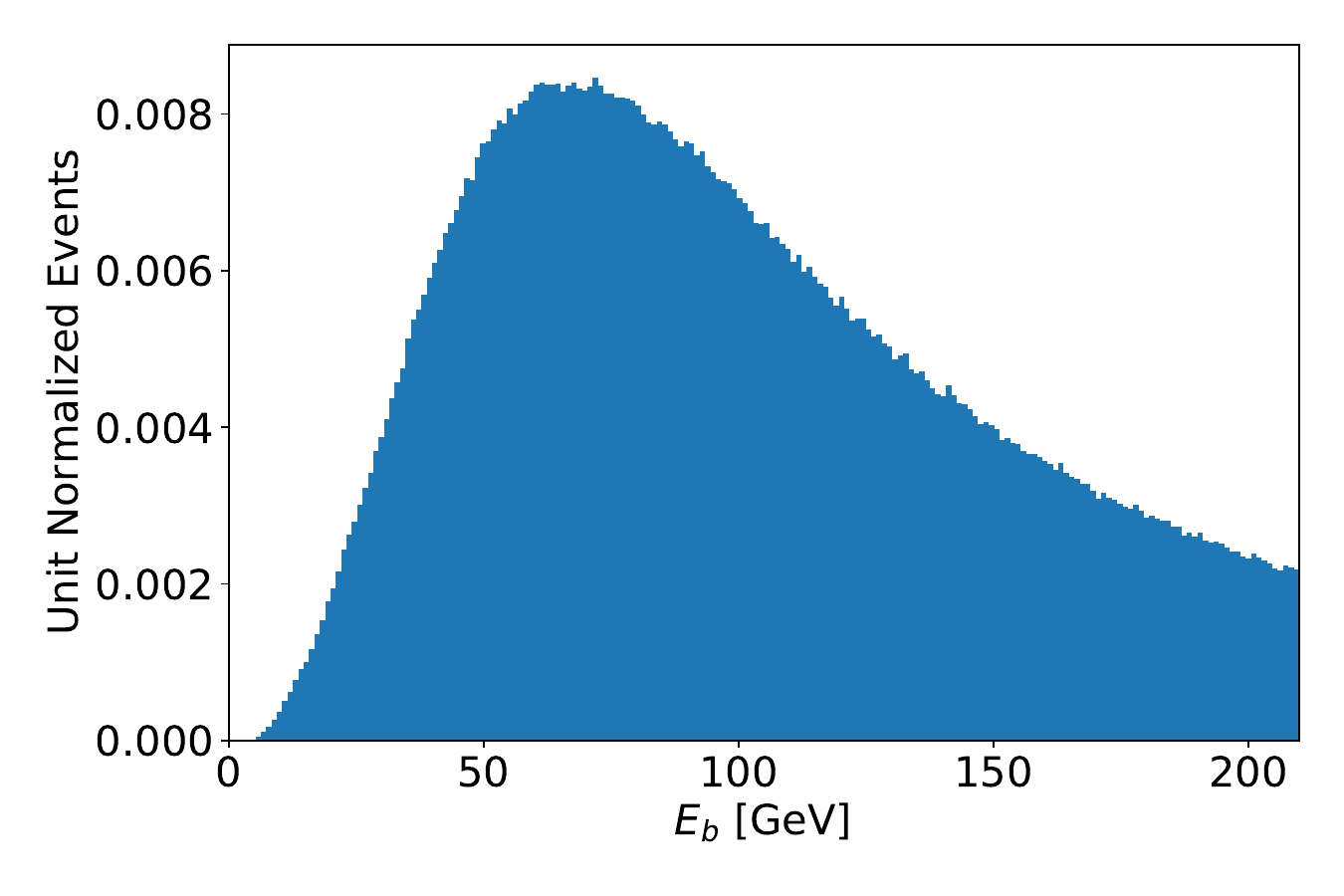}
    \caption{Energy spectrum of the bottom quark. }
    \label{fig:EbfromTop}
\end{figure}

Next, we point out that our peculiar choice of observable for the determination of the top quark mass is associated with an energy peak that is rather {\it broad}, as shown in Figure~\ref{fig:EbfromTop}. To reliably extract the peak position from the data, it is thus important to minimize any potential distortion of the distribution in the vicinity of the peak. As we will delineate in the next section, our event selection scheme is designed to not only sufficiently suppress the backgrounds but also accommodate this requirement. 
Indeed, a simple analytic {\it fitting} function for the energy distribution has been proposed in Ref.~\cite{Agashe:2012bn}:
\begin{eqnarray}
f^{ \rm fit } \left( E_b; E^{ \rm rest }_b, w \right) & = & \frac{1}{ N } \exp \Big[ - w  \left( \frac{ E_b }{ E_b^{ \rm rest } } +  
\frac{ E_b^{ \rm rest } }{ E_b } \right) \Big]\,,
\label{ansatz_Eb}
\end{eqnarray}
where $w$ and $E_b^{\rm rest}$ are the fitting parameters and $N$ is the overall normalization factor. The $w$ parameter describes the width of the distribution, while $E_b^{\rm rest}$ takes care of the peak position, and its best-fit value is equated to the rest-frame energy value in Eq.~\eqref{Ebrest}.

This was inspired 
by properties of the energy distribution, in addition to the peak position, that can be proven from first principles\footnote{For example, we can show that the energy distribution {\em must} be a function of the combination 
$E_b^{\rm lab} / E_b^{\rm rest}$.} and checked against a
parton-level inclusive calculation, (see LHS of Fig.~1 of Ref.~\cite{Agashe:2012bn}).\footnote{Note that the CMS implementation used a different fitting function, essentially a Gaussian in
{\em logarithm} of $E_b$ (instead of the above form). The form adopted by CMS still has the expected properties
reported in Ref.~\cite{Agashe:2012bn}.}
Finally, this fitting procedure was shown to be robust against realistic effects such as the application of selection criteria and smearing of measured quantities due to detector effects as implemented in \texttt{Delphes}~\cite{de-Favereau:2013dk}, (see RHS of Fig.~1 of Ref.~\cite{Agashe:2012bn}).
For these reasons we will define our peak-energy proxy and evaluate the goodness of our top mass extraction based upon an analysis that uses a template from the family in Eq.~(\ref{ansatz_Eb}), or possible extensions of this family based on arguments presented in Ref.~\cite{Agashe:2012bn}.

\subsection{Event generation and selection criteria for $B$-hadron decay length}

For the final states in Eqs.~(\ref{eq:ttbar2emu})-(\ref{eq:ttbar2eemumu:semileptonic}) we consider the processes 
\begin{eqnarray}
    p p & \rightarrow & t \bar{t} \ (t \rightarrow W^+ b, \ W^+ \rightarrow \nu_l l^+), \ (\bar{t} \rightarrow W^- \bar{b}, \ W^- \rightarrow \bar{\nu_l} l^-)\,,\\
     p p & \rightarrow & t \bar{t} \ (t \rightarrow W^+ b, \ W^+ \rightarrow ff^{'}),  (\bar{t} \rightarrow W^- \bar{b}, \ W^- \rightarrow \bar{\nu_l} l^-) + c.c.\,.
\end{eqnarray}  
 We use \texttt{MadGraph5\_aMC@NLO}~\cite{Alwall:2014hca} as the parton-level Monte Carlo event generator for proton-proton collisions to generate $t \bar{t}$ events, specifying their decay chains to be either semi-leptonic or dileptonic with the center-of-mass energy being $\sqrt{s} = 14 \text{ TeV}$. First, we generate fully inclusive events, and then we impose event selection criteria similar to what the CMS and ATLAS Collaborations use to eliminate background events~\cite{CMS:2016iru}. 
 After some further optimization for semi-leptonic events, described below, we require either one electron with $p_T>25 \text{ GeV}$ and $|\eta|<2.4$ or one muon with $p_T>25 \text{ GeV}$ and $|\eta|<2.1$. We also require at least 4 jets with $p_T>25 \text{ GeV}$ and $|\eta|<2.5$.\footnote{A residual JES dependence associated with event selection criteria that makes use of calorimeter-based jet energies was avoided via the use of track-based jets in Ref.~\cite{CDF:2009qxf}.} 
 For same-flavor dilepton events, as a baseline, we require exactly two leptons with $p_T>25 \text{ GeV}$ and $|\eta|<2.4$, at least 2 jets with $p_T>25 \text{ GeV}$ and $|\eta|<2.5$, $E_T^{\rm miss}> 40 \text{ GeV}$, and for the invariant mass of the two leptons, $M_{\ell\ell}$, we require   $|M_{\ell\ell} - m_{Z}| > 15 \text{ GeV}$ to remove backgrounds containing a real $Z$ boson and  $M_{\ell\ell} > 20 \text{ GeV}$ to remove backgrounds from photon conversions and QCD resonances. In Section~\ref{results} we show the effect of variations of these optimal selection criteria.\footnote{The proposed cuts we show are similar to those used by experiments to eliminate background events. We expect our minimal changes not to affect background estimates on identifying $t\bar{t}$ events.} For comparison, the CMS selection for a determination of $m_{t}$ using a $B$-hadron decay length measurement are reported in Table~\ref{tab:baseline-and-optimal}, together with a summary of our selection criteria. Our selection is identical for the $\eta$ ranges but differs on the $p_T$ requirements. As detailed later, our analysis prefers to have equal $p_T$ thresholds for all final states, hence we softened the jet $p_T$  and hardened the lepton $p_T$ requirements to a medium value of 25~GeV, which should be attainable for experiments at the LHC and HL-LHC.

\begin{table}
\centering
\resizebox{\columnwidth}{!}{
\begin{tabular}{|c|c|c|c|}
\hline 
 &  &   Ref.~\cite{CMS:2016iru} & Optimal choice for our analysis\tabularnewline
\hline 
\hline 
\multirow{3}{*}{$\ell+{\rm jets}$} & $e$ & $p_{T}>30$~GeV, $\eta<2.4$ & $p_{T}>25$~GeV, $\eta<2.4$\tabularnewline
\cline{2-4} \cline{3-4} \cline{4-4} 
 & $\mu$ & $p_{T}>26$~GeV, $\eta<2.1$ & $p_{T}>25$~GeV, $\eta<2.1$ \tabularnewline
\cline{2-4} \cline{3-4} \cline{4-4} 
 & $j$ & $N_{j}\geq4$, $p_{T}>30$~GeV, $\eta<2.5$ &$N_{j}\geq4$, $p_{T}>25$~GeV, $\eta<2.5$  \tabularnewline
\hline \hline
\multirow{5}{*}{$2\ell+{\rm jets}$} & $e,\mu$ & $p_{T}>20$~GeV, $\eta<2.4$ & $p_{T}>25$~GeV, $\eta<2.4$ \tabularnewline
\cline{2-4} \cline{3-4} \cline{4-4} 
 & SF & $M_{\ell\ell}>20$~GeV, $|M_{\ell\ell}-m_{Z}|>15$~GeV & $M_{\ell\ell}>20$~GeV, $|M_{\ell\ell}-m_{Z}|>15$~GeV \tabularnewline
\cline{2-4} \cline{3-4} \cline{4-4} 
 & OF & - & - \tabularnewline
\cline{2-4} \cline{3-4} \cline{4-4} 
 & $j$ & $p_{T}>30$~GeV, $\eta<2.5$ & $p_{T}>25$~GeV, $\eta<2.5$ \tabularnewline
\cline{2-4} \cline{3-4} \cline{4-4} 
 &  & $E_T^{\rm miss}>40$~GeV & $E_T^{\rm miss}>40$~GeV \tabularnewline
\hline \hline
\end{tabular}
}
\caption{Baseline selection of the events used in our analysis and an optimized choice that we use to minimize bias of the measured top quark mass.\label{tab:baseline-and-optimal}}
\end{table}

The events that have passed our selection cuts are fed into \texttt{Pythia~8.2}~\cite{Sjostrand:2014zea} for showering and hadronization. 
 The $B$-hadron decay lengths are then fitted to the double convolution function to extract the $b$ quark energy peak as per Eq.~(\ref{ansatz_LB}). 

\subsection{Extracting energy peak from $B$-hadron decay lengths}
\label{practicaldecay}

For a realistic discussion, we need to take into account some crucial differences between $b$-jets and $B$-hadrons. 
The most obvious is the fact that a bottom quark can hadronize into one of the many $B$ hadron species, with different masses and mean rest-frame lifetimes. Therefore, constructing a fitting function for $B$-hadron energy or length spectrum is unavoidably more involved than what we sketched in Eq.~(\ref{double-convolution}). The probability of decay into a particular $B$ hadron species needs to be taken into account, thus we introduce fractions $f_i$ for the production yield of each hadron $B_i$, whose mass, energy, and lifetime need to be correctly tracked in our formula.  This leads to the following more appropriate fitting function for the spectra:

\bea
G^{ \rm fit } \left( L_B; E^{ \rm rest }_b, w ,\nu \right) & = & \int_{ E_{B,\rm min} }^{E_b} d E_B \int_{E_{b,\rm min}}^{E_{b,\rm max}} d E_b  \frac{1}{ N (w) } \exp \bigg[ - w  \left( \frac{ E_b }{ E_b^{\rm rest} } +  
\frac{ E_b^{\rm  rest } }{ E_b } \right)^{\nu} \bigg]  \times \nonumber  \\
& & \nonumber
\sum_i D_i \left( \frac{ E_{B_i} }{ E_b }; E_b \right) 
\frac{ f_i m_{B_i} }{ c \tau_{B_i}^{ \rm rest }\sqrt{E_B^2 - m_{B_i}^2 } }
\exp \left( - \frac{ L_B m_{B_i} }{ c \tau_{B_i}^{ \rm rest } \sqrt{E_B^2 - m_{B_i}^2} } \right),\nonumber \\
\label{ansatz_LB}
\eea
\noindent where the functions, variables, and parameters in this expression play the following roles:
\bea
G^{ \rm fit } \left( L_B; E^{ \rm rest }_b, w \right) & \rightarrow & \hbox{fitting function for the decay length $(L_B)$ distribution} \nonumber \\
\hbox{best-fit value of parameter} \; E^{ \rm rest }_b & \rightarrow & \frac{ m_t^2 - M_W^2 + m_b^2 }{ 2 \; m_t } \nonumber \\
\tau_B^{ \rm rest } & \rightarrow & \hbox{mean decay lifetime of $B$-hadron in its rest frame} \nonumber \\
\hbox{parameter} \; w & \rightarrow & \hbox{width of fitting function} \nonumber \\
\hbox{parameter} \; \nu & \rightarrow & \hbox{tails weight of fitting function} \nonumber \\
i & \rightarrow & B\text{-hadron species} \nonumber \\
D_i \left( \frac{ E_{B_i} }{ E_b }; E_b \right) & \rightarrow & \hbox{bottom quark fragmentation function for species } i \nonumber \\
f_i & \rightarrow & \text{relative proportion of species } i \nonumber \\
N ( w ) & \rightarrow & \hbox{normalization factor} 
\nonumber
\eea

Concerning hadronization, we note that the relative fractions $f_i$, in principle, depend on the kinematics of the bottom quark, see e.g. \cite{LHCb:2014ofc,HFLAV:2019otj,LHCb:2013vfg,ATLAS:2015esn} for the $p_T$ and $\eta$ dependence of $\Lambda_b$ and $B_s$ production, and therefore on the production mechanism, including a subtle dependence on the color environment, which raises some doubt about the universal applicability of results from $e^+e^-$ machines to hadron machines at the level of precision that our method may require.

In order to disentangle the role of uncertainty in $D_i$, discussed in detail below in Sec.~\ref{sec:hadronization}, from other uncertainties in the method, we largely eliminate it here by computing our templates using exactly the same $D_i$ as in the generation of data with Pythia. In particular, for all $b$ jets in our simulation sample, we extract the energies of the $b$ quarks before showering and the energies of the subsequently formed $B$-hadron to obtain the distribution of $D_i(\frac{E_{B_i}}{E_b};E_b)$ for each hadron species $B_i$ in Pythia.

We stress that by measuring this fragmentation function from Pythia itself and plugging it into our double integral, we essentially remove the unknowns due to fragmentation in our procedure. Therefore, the results we will obtain in this way can be considered to be the best achievable results for negligible uncertainties originating from fragmentation. We will elaborate further below about the possible impact of uncertainties in fragmentation and hadronization. 
{We remark that our $D_i$ functions do have a dependency on $E_b$. This dependence arises in part from the fact that some showering physics is encapsulated in these functions. We verified that the $E_b$ dependence is quite weak and so in the following we will drop it to keep our exercise computationally manageable.}
In the following, we calculate the sensitivity of the top mass determination to knowledge about the relative fractions and fragmentation functions in order to estimate the uncertainties in the final result that should be associated with them. Results on these matters are presented in Section~\ref{results}.

To compute the templates of Eq.~(\ref{ansatz_LB}), we need to fix some parameters. Concerning the integrals, we set the lower and upper limits of the $E_b$ integration range to 40 GeV and 450 GeV, respectively.
For the energy $E_B$ we impose a minimum energy requirement, $E_B>7$~GeV
and a fragmentation boundary constraint $E_B<E_b$.
These range choices were chosen to minimize both error and bias in our result. Variations of these ranges have been tested and are discussed in Section~\ref{results}.

For our top quark mass extraction, we will consider only a limited range of $L_B$, to maximize the stability of our results and the precision attainable. We found that the calculation of the double convolution becomes numerically challenging when evaluated for very large decay lengths. As there {is only 5\% of the total data  at $L_B>20$~mm}, we use the range $L_{B}\in [0,20]$~mm. For our measurement, there is no lower limit on the range of $B$-hadron decay lengths, but in practice, experiments have a minimum threshold of $\mathcal{O}(1)$~mm on decay length. 
We find that our results change negligibly when we increase the lower limit of the $B$-hadron decay length to 1 mm. As this threshold is experiment-dependent, we quote results for a fitting range starting at 0 mm in the following.

The exponent $\nu$ in Eq.~(\ref{ansatz_LB}) is in principle a parameter of our template that would need to be floated in the fit. This is a new parameter compared to our original ansatz of Ref.~\cite{Agashe:2012bn}, which corresponds to $\nu=1$. Variations of $\nu$ mostly change the shape of the $b$ quark energy spectrum in the tail regions, which have a greater importance for this application than in our earlier work. Its value participates in the definition of our template and we find $\nu=0.3$ as optimal for minimizing overall bias in our result. In principle, $\nu$ can be measured independently of our method by fitting the tails of the $b$-jet energy spectrum. We measured $\nu = 0.3$ by fitting the tails of Monte Carlo generated $b$-jet energies. We note that the bias our method is not very sensitive to our choice of $\nu$.

\begin{table}
\begin{centering}
\begin{tabular}{|c|c|}
\hline 
 & best   \tabularnewline
\hline 
\hline 
$\nu$  & 0.3  \tabularnewline
\hline
$E_{b}$ range & {[}40,450{]} GeV   \tabularnewline
\hline 
$E_{B}$ & $7{\,\rm GeV}<E_{B}<E_{b}$  \tabularnewline
\hline 
$L_B$ &  [0,20]~mm   \tabularnewline
\hline 
\end{tabular}
\par\end{centering}
\caption{\label{tab:hyper-parameters-best} Summary of the parameters that we fixed to compute our template Eq.~(\ref{ansatz_LB}).}
\end{table}

Having set these hyper-parameters, as summarized in Table~\ref{tab:hyper-parameters-best}, all that remains to be done is to fit the decay length data to the fitting function Eq.~(\ref{ansatz_LB}), and obtain $E_b^{\text{rest}}$ and its uncertainty from a $\chi^2$ analysis. Then $m_{\text{top}}$ is obtained from $E_b^{\text{rest}}$. 

Ideally, we want to have a precise measurement, that stems from a narrow $\chi^2$, with no bias from our template fits. In other words, the minimum of the $\chi^2$ should coincide with the input top quark mass. 
We estimate the bias by doing pseudo-experiments for various combinations of the hyper-parameters and we pick the values which lead to the smallest bias. These optimal values are collected in Table~\ref{tab:hyper-parameters-best}. The stability of the estimated bias and details on the whole procedure are given in Section~\ref{results} together with our main result.

\section{Results \label{results}}
\subsection{Measurement of the top quark mass}

\begin{figure}
    \centering
    \includegraphics[width = 0.59\textwidth, trim = {0 0 0 40}, clip]{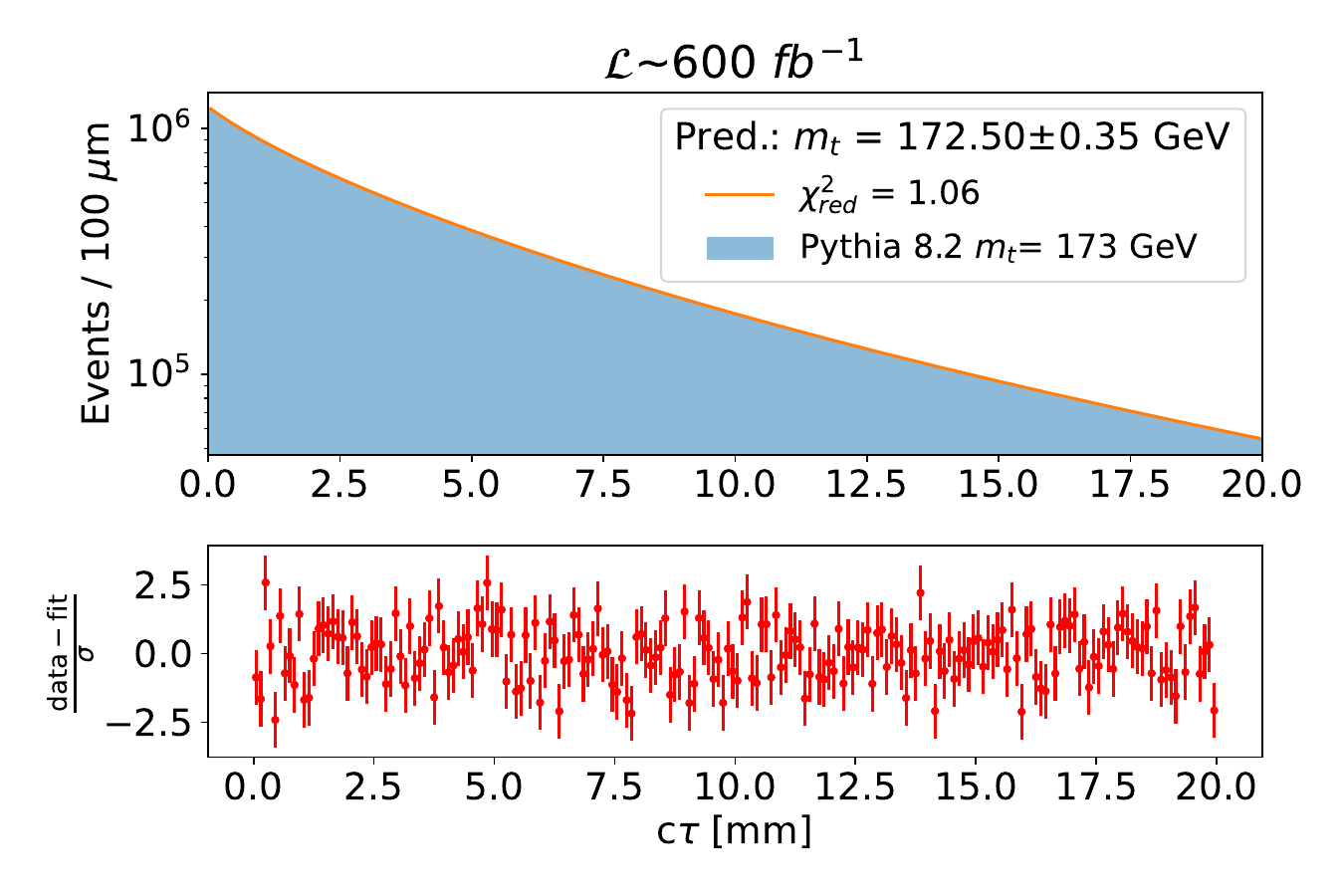}
    \caption{Pythia 8.2 decay length spectrum for input $m_t = 173$ GeV, normalized for $\sim 230$/fb at LHC 14 TeV summing fully leptonic and semileptonic $t\bar{t}$ decays (blue). Result of the fit  not corrected for the bias (orange).}
    \label{fig:sample_fit}
\end{figure}

\begin{figure}
    \centering
    \includegraphics[width = 0.49\textwidth]{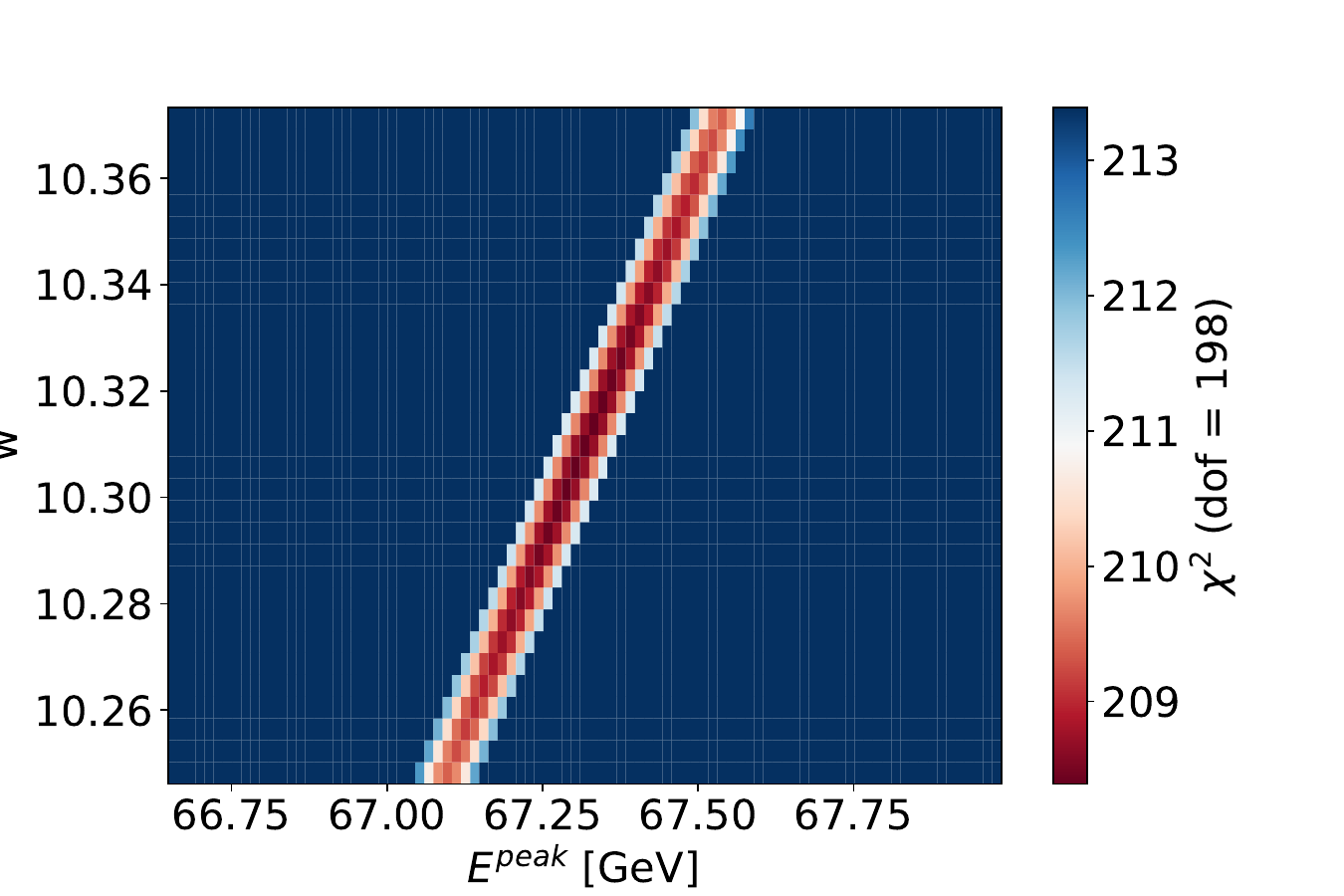}
      \includegraphics[width = 0.49\textwidth]{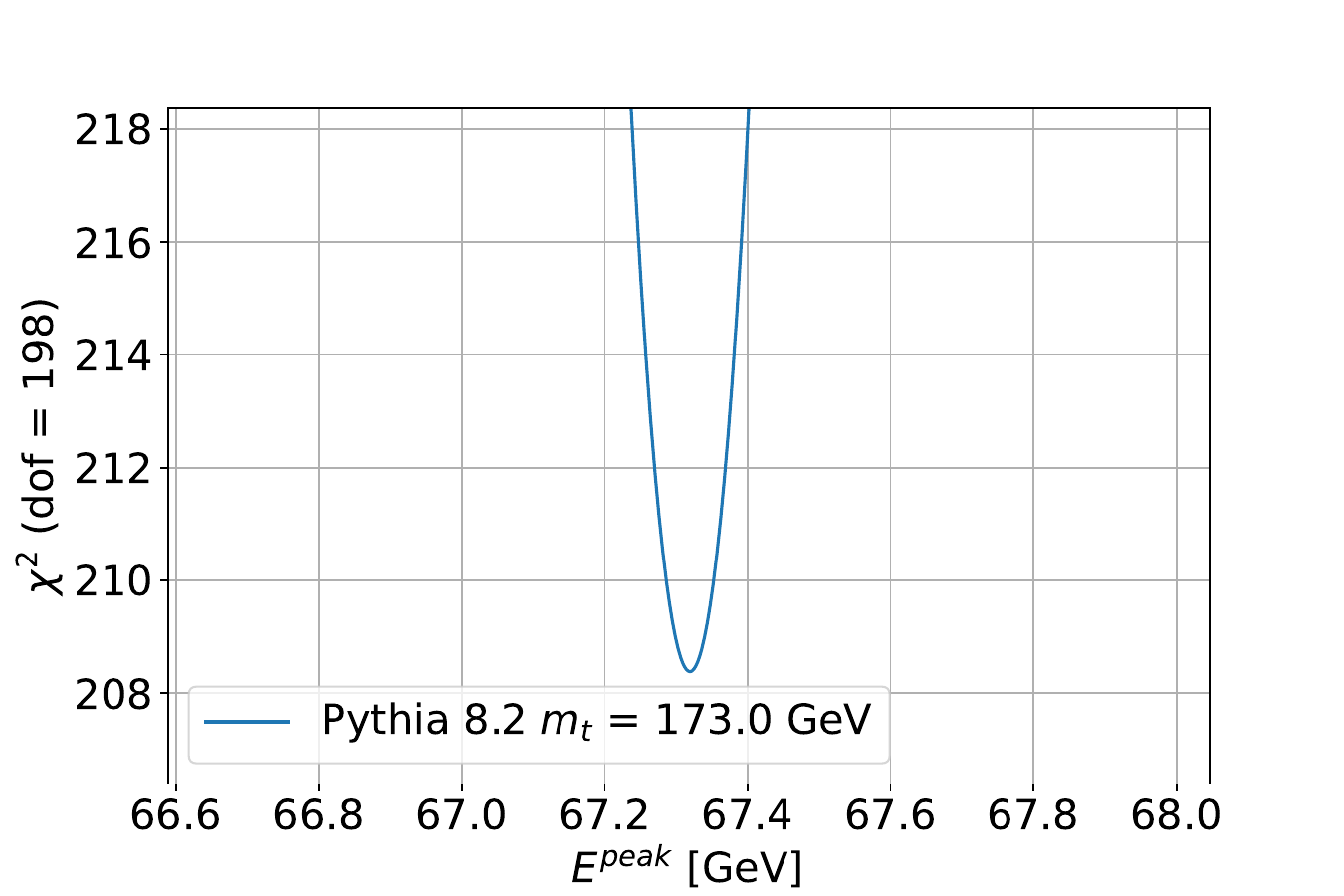}
    \caption{$\chi^2$ profile in the vicinity of the best-fit parameters in the space $(E^{\rm rest}_{b}$, $w)$ (left) and for fixed $w$ as a function of $E_b^{\rm rest}$ (right).  }
    \label{fig:chi_profile}
\end{figure}

The measurement of the top quark mass starts from the observed decay length spectrum, inclusive of all species of $B$-hadrons that are produced at the LHC in $\ttbar$ events. A prediction from this observable from Pythia~8.2 is shown in Figure~\ref{fig:sample_fit} for $m_t=173.0\gev$. Using our template Eq.~(\ref{ansatz_LB}) we can fit the best value for $m_t$ with a simple $\chi^2$ minimization and find $m_t=172.50\pm 0.35$ on this particular spectrum. An example $\chi^{2}$ shape is displayed in Figure~\ref{fig:chi_profile} for illustration.

{The  result shown in Figure~\ref{fig:sample_fit} is just one instance of a representative input mass; $m_{t\text{,~input}}=173.0$~GeV. In the following, we present the results obtained for various other masses and we quote the performance of our method by taking the average of the uncertainties obtained. In particular, we apply our template fitting procedure to top quark masses over the range $170 \text{ GeV} \leq m_{t\text{,~input}} \leq 176 \text{ GeV}$ with the hyper-parameters fixed to the values in Table~\ref{tab:hyper-parameters-best}} to obtain the following expression for the uncertainty, which is discussed in detail below:
\beq
    \delta m_t^{(E_B,~peak)} = \frac{0.5~\rm GeV}{\sqrt{\mathcal{L}/100\,{\rm fb}^{-1}}} ({\rm stat}.) 
\oplus  0.5~{\rm GeV} \cdot \left( \frac{0.1\%}{\frac{\delta{D_{i}}}{D_{i}}}\right) \;
\oplus  0.3~{\rm GeV}\cdot \left(  \frac{5\%}{\frac{\delta{f_{i}}}{f_{i}}} \right) \;
. \label{eq:money}
\eeq

Given the need to compute templates, our method incurs a risk of returning a value of the top quark mass that has a small uncertainty, due to a sharp $\chi^2$ profile, but is somewhat shifted with respect to the input value. As the event generators currently available in QCD are not fully up to the task of dealing with the $10^{-3}$ precision that a top quark mass measurement currently demands, we take the position that a robust method for the measurement of the top quark mass should be designed with the goal of a small bias as well as a small error bar. In fact, lacking a reliable event generator for the precision we need, this approach has the potential to avoid a large bias when applied to real-world data, which will certainly differ from the Monte Carlo data on which the bias and the error bar have been optimized.
When we set all the parameters to their optimum values \sa{by tuning them}, we obtain an average bias in the $m_t$ measurement that is  $-0.18 \pm 0.12 $ GeV, which is below the expected systematic uncertainty. It is also not very sensitive to the choice of hyper-parameters, as detailed below Section~\ref{sec:templateUncertainties}, thus we consider our method to have negligible bias. 

With an integrated luminosity of $100~{\rm fb}^{-1}$, we find that  at 14~TeV LHC the expected statistical uncertainty for a $\chi^2$ fit of our templates, assuming a total production cross-section for $t\bar{t}$ at 985~pb~\cite{Czakon:2013goa}, is $500$~MeV.
Thus we expect this measurement to be already competitive with regard to the statistical uncertainty over 
other historically more frequently pursued methods in LHC Run~3. We further expect that the statistical uncertainty will go down to about 100~MeV at HL-LHC for 3000~$\rm{fb}^{-1}$. Hence, a thorough discussion of the systematic uncertainties associated with this method is now in order. 

\subsection{Uncertainty from template definition\label{sec:templateUncertainties}}
    \begin{figure}
        \centering
        \includegraphics[scale = 0.7, trim = 0 0 0 30, clip]{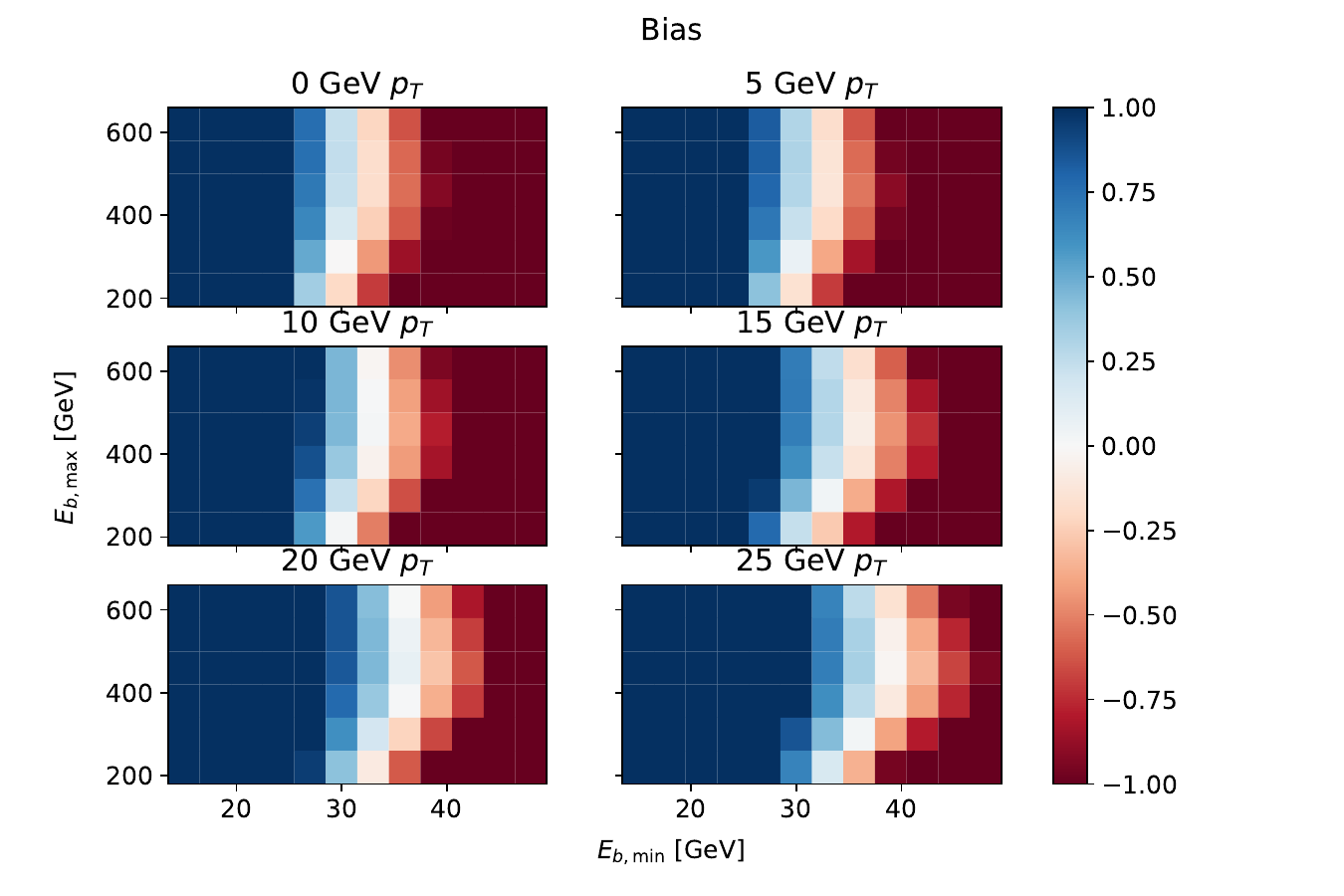}
        \caption{Bias as a function of the limits on the $b$-quark energy range in the $E_b$ integral of Eq.~(\ref{ansatz_LB}). Subplots are titled by the common $p_T$ cut on leptons and jets used for the selection of events. 
       }
        \label{fig:average_bias}
    \end{figure}
    
    \begin{figure}
        \centering
        \includegraphics[width=0.49\textwidth, trim = 0 0 0 30, clip]{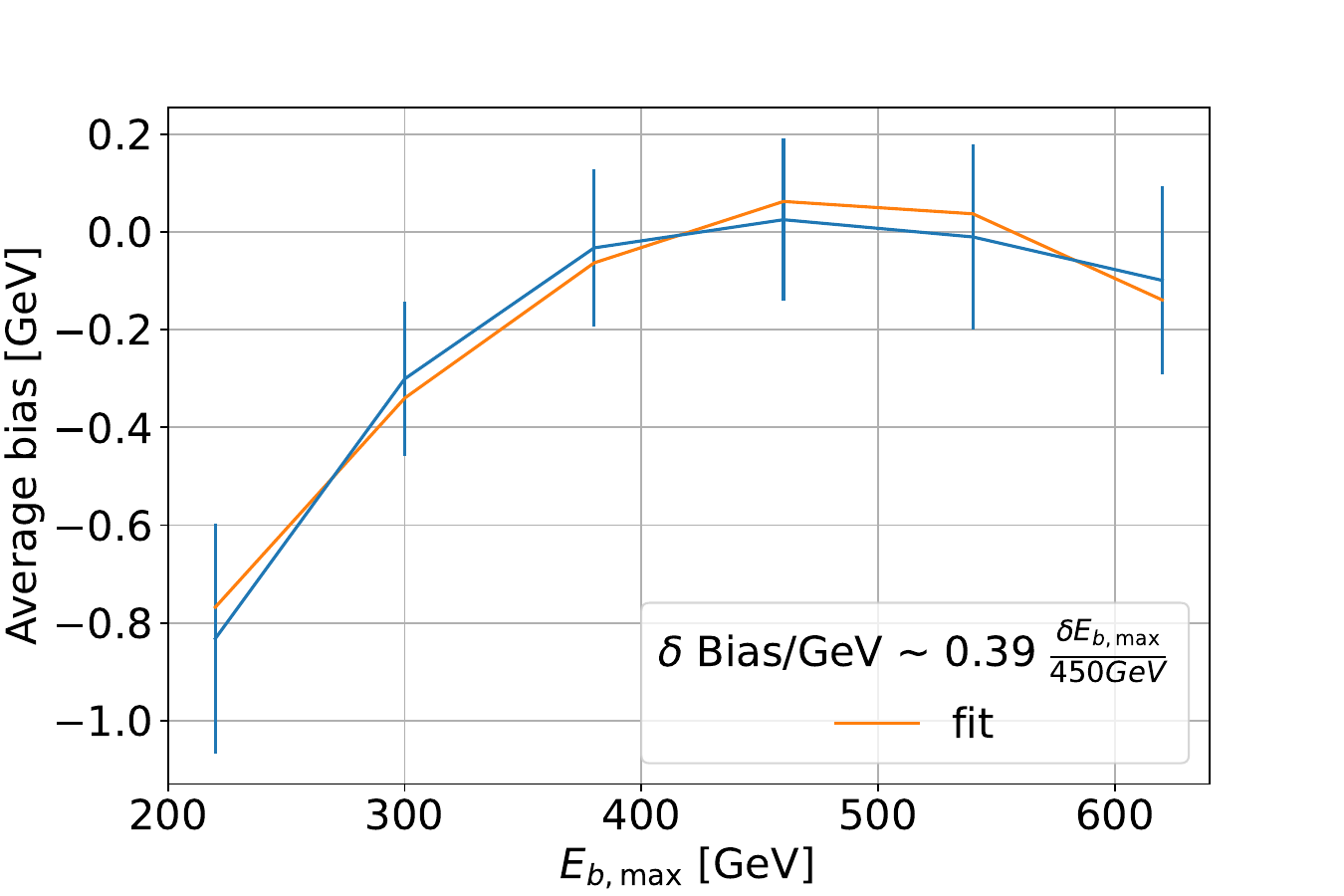}
            \includegraphics[width=0.49\textwidth, trim = 0 0 0 30, clip]{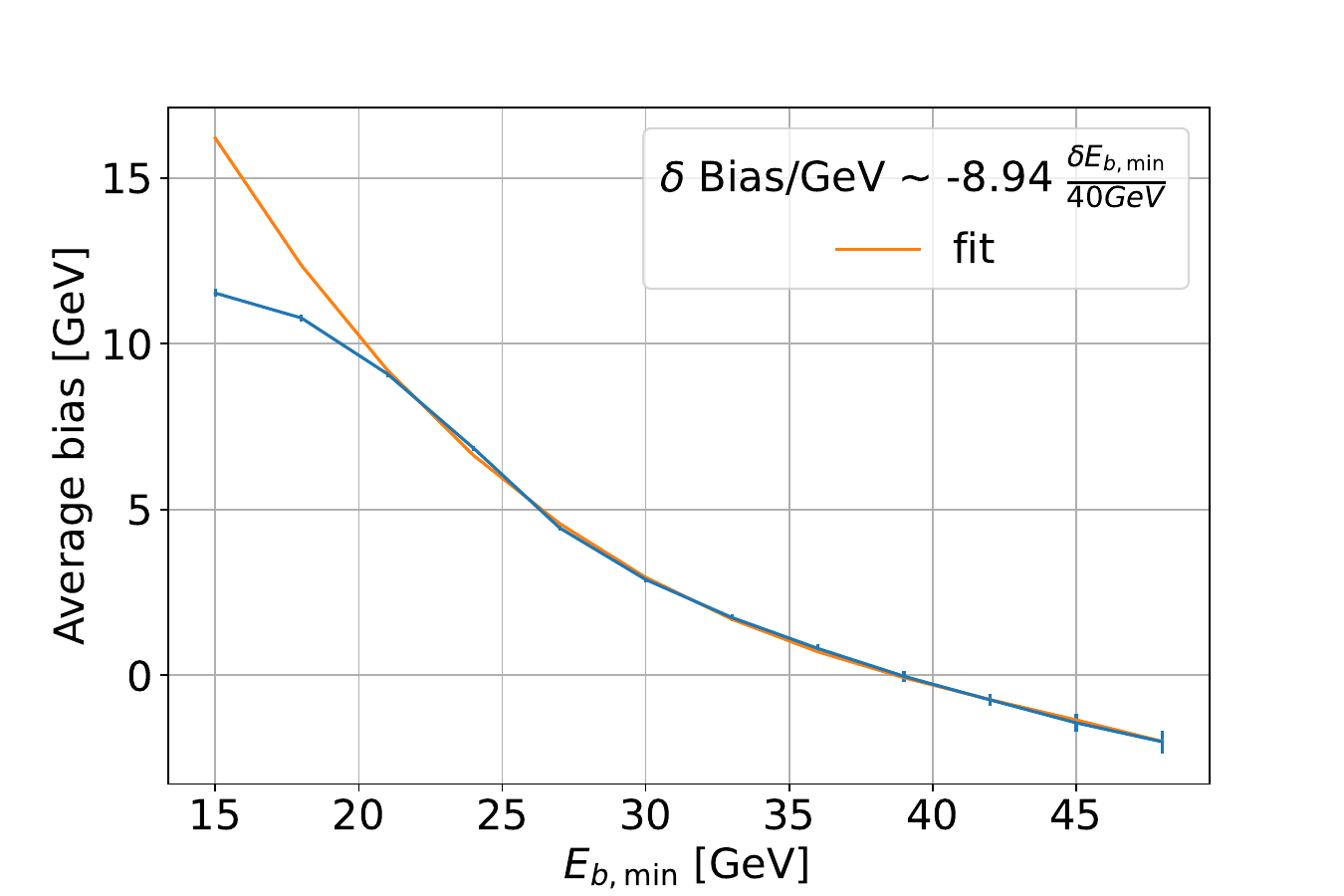}
            \caption{Bias as a function of the lower and higher limit on the b-quark energy range on which we perform the $E_b$ integral in Eq.~(\ref{ansatz_LB}). The lower (higher) limit is fixed at 450 GeV (40 GeV) in the right (left) plot.  }
            \includegraphics[width = 0.49\textwidth, trim = 0 0 0 30, clip]{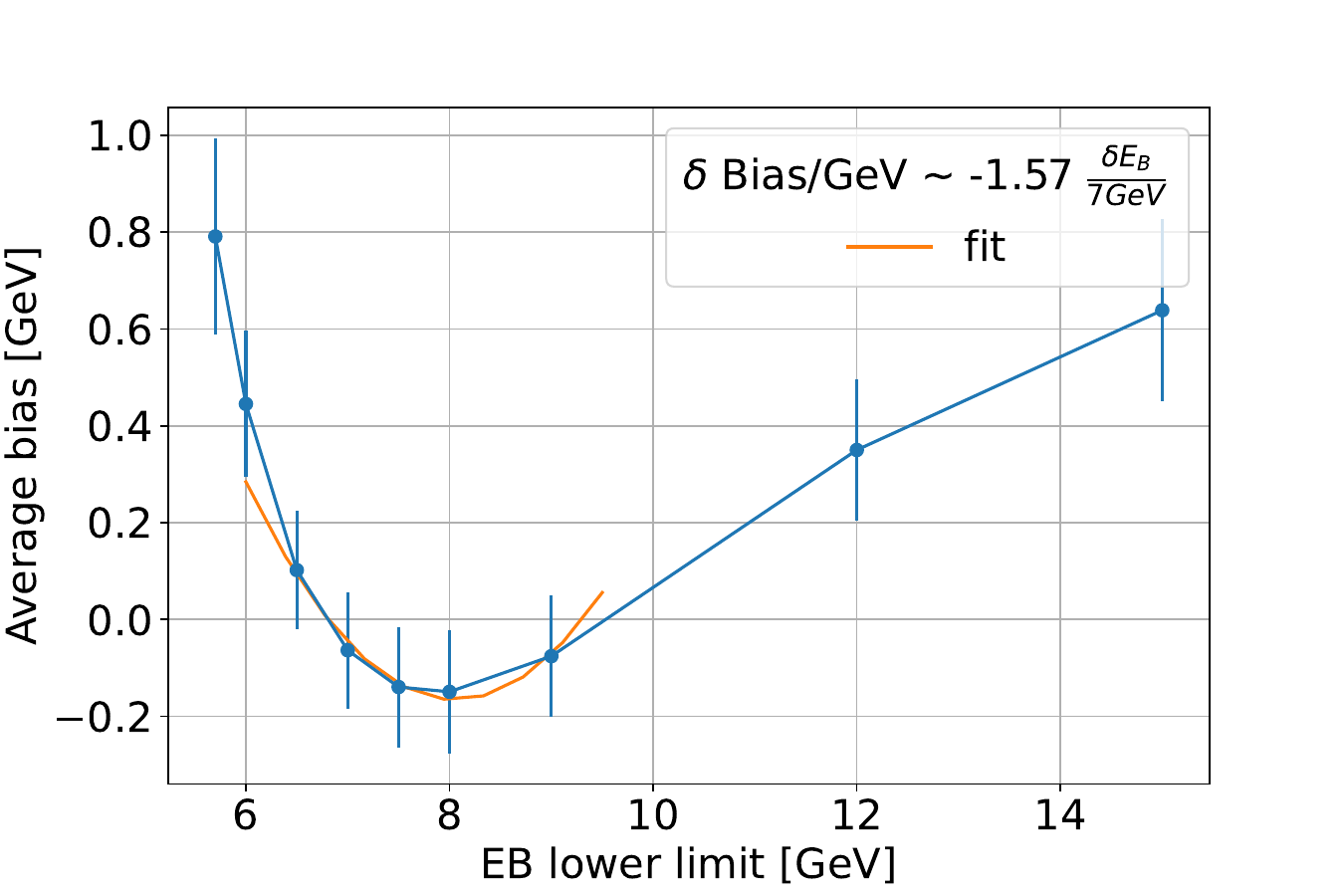}
            
            \caption{The variation of the bias as a function of the cut on minimum B-hadron energy. The lower and the higher limit on b-quark energy is fixed at 40 GeV and 450 GeV respectively, as in Table~\ref{tab:hyper-parameters-best}.}
        \label{fig:average_bias_1D_profiles}
    \end{figure}
    
    The definition of our template Eq.~(\ref{ansatz_LB}) requires that some parameters be fixed and a phase-space region be specified in which we want to carry out the measurement as summarized for our analysis in Tables~\ref{tab:baseline-and-optimal} and \ref{tab:hyper-parameters-best}. As noted above, the goal for these choices was a minimization of both the bias and the error in our measurement. Variations of these choices need to be explored to demonstrate the stability of the procedure. It is important to verify that it does not hinge on a too particular or peculiar a choice of the phase-space region or a too narrow definition of the template parameters. In fact, a mismatch between the phase-space used to compute the templates and the phase-space imposed on the data through event selection criteria may be reflected in a mismatch between the extracted top quark mass and its true value.
    
    We quantify this possible mismatch with a study of the impact of variations of the $p_T$ cut on each particle and of the $E_b$ range in our template definition. For this study, we fix the choice of the length fitting range $L_B \in [0, 20] \text{ mm}$ as per our optimal measurement strategy. We find that using unequal $p_T$ cuts for different particles tends to introduce larger biases, hence in the following, we consider a common value for the $p_T$ cut of both leptons and jets. We vary  the common $p_T$ cut on jets and leptons in the event selection criteria and   the limits on bottom quark energy and we obtain the bias (in GeV) on the top quark mass shown in Figure~\ref{fig:average_bias}. 
    We found that the choice of the $L_B$ fitting range is not very important as long as the tail of the spectrum is excluded.
    All in all, an analysis can be carried out with a common $p_T$ threshold at  25~GeV in the energy range $E_b \in [40,450]$~GeV. As shown broadly in Figure~\ref{fig:average_bias} and detailed in Figure~\ref{fig:average_bias_1D_profiles}, a variation of the highest energy considered in the template would have a very mild effect on the bias. Variations of the lowest energy considered in the template may have more impact, however, the bias is around $\frac{ 80{\rm MeV}}{ \left(\frac{\delta E}{E}/10^{-2}\right)}$, which is reasonable to neglect in view of the expected  $\delta E/E$ performance for $b$-jets in Run3 and HL-LHC experiments. Variations of the minimum $B$-hadron energy can lead to bias of the extracted $m_t$ by around 160~MeV for a 10\% error on $E_B$. Experimental analysis will have to select $B$-hadron decay modes for which such precision on the measurement of $E_B$ can be attained.

  \subsection{Hadronization and fragmentation uncertainties \label{sec:hadronization}}
    In our analysis, we use multiple $B$-hadron species, which differ in mass, mean decay length, and production rate. The central values that we used for these quantities are given in Table~\ref{tab:B_hadron_prop}. To estimate the sensitivity of the result to these inputs in the ansatz, we vary them one by one  and see how the prediction for the extracted top quark mass shifts on a fixed data set. Clearly, future measurements will be needed to better fix  these quantities, especially if they can be measured directly at the LHC in phase-space regions similar to $b$ quarks from $\ttbar$ production.
       
    \begin{table}[]
        \centering
        \begin{tabular}{|c|c|c|c|}
        \hline \hline
            Hadron & Mass (MeV)~\cite{ParticleDataGroup:2020ssz} & Lifetime ($10^{-12} $ s)~\cite{HFLAV:2019otj} & Fraction \\
            \hline
             $B^{\pm}$ & 5279.34 $\pm$ 0.12  & 1.638 $\pm$ 0.004 & 42.9 \%\\
             $B^0$ & 5279.65 $\pm$ 0.12  & 1.519 $\pm$ 0.004 & 42.9 \%\\
             $B_{s}^0$ & 5366.88 $\pm$ 0.14  & 1.516 $\pm$ 0.006 & 9.5 \%\\
             $\Lambda_b^0$ & 5619.69 $\pm$ 0.17  & 1.471 $\pm$ 0.009 & 3.6 \%\\
             \hline \hline
        \end{tabular}
        \caption{Properties of the four most prominent species of $B$ hadrons from $b$-quark hadronization. Production fractions are taken from Pythia~8.2 Monash tune default.}
        \label{tab:B_hadron_prop}
    \end{table}
    \begin{table}[]
        \centering
        \begin{tabular}{|c|c|}
    
            \hline \hline
            Parameter 
            & Sensitivity \\
            \hline
            $m_{B_i}$ 
            & $\simeq 1$ \\
            \hline
            $\tau_{B_i}^{\text{rest}}$ 
            & $\lesssim 1$ \\
            \hline
            $f_i$ 
            & $\simeq 0.04 $ \\
            \hline \hline
        \end{tabular}
        \caption{Sensitivity of the top quark mass measurement to the properties of B hadron species involved. The sensitivity that we quote is the maximum sensitivity across the hadron species.}
        \label{tab:Sensitivity}
    \end{table}
    
Table~\ref{tab:Sensitivity} summarizes how the extracted top quark mass changes upon variations of  $B$ hadrons properties. We quote a sensitivity
\beq 
\Delta_{\xi_{B_j}} = \frac{ {\delta m_t \over m_t}}{{\delta \xi_{B_j} \over \xi_{B_j}} } \label{eq:sensitivity-ratio}
\eeq 
computed as the ratio of the relative variation on the extracted $m_t$ over the variation of the parameter $\xi_{B_j}$ for each $B$ hadron property $\xi_{B_j}$ listed in Table~\ref{tab:B_hadron_prop}. The shift in the extracted mass tends to be bigger when properties of $B^{\pm}$ or $B^0$ hadrons are varied because these species have a higher production fraction in $b$ quark fragmentation. In the right column, we quote the maximum shift in $m_t$ that we get among the four hadron species considered. The result is that for masses and lifetimes we need to know the physical values with the same or better precision that we want to extract the top quark mass. The present knowledge of these quantities seems more than adequate to warrant the resulting uncertainty on the top quark mass to be negligible.  

Production fractions tend to impact the extracted top quark mass to a much milder degree than masses and lifetimes, hence relatively large uncertainties up to 5\% or so can be tolerated and still give rise to sub-leading uncertainties in the extracted top quark mass. We ascribe this result to the fact that the properties of different $B$-hadron species are rather similar. The universality of the presently measured values of $f_i$ is questionable when different production environments and different kinematic regimes are compared, so we cannot really use the uncertainty of present measurements at LEP and the TeVatron for the $f_i$ as a guideline for the impact of $f_i$ uncertainty on the extraction of $m_t$. In any case, it seems that if the LHC will attain even just the same $\mathcal{O}(10\%)$ TeVatron precision on the measurement of $f_i$, this uncertainty will have an impact on the top mass extraction of order 100~MeV at most.  It should be noted that the fractions $f_i$ have been proven to be functions of the transverse momentum of the hadron (see e.g. Section 4.1.3 of \cite{HFLAV:2019otj} for a recent summary). However, this effect should be of even more sub-leading importance for the $f_i$ values, hence it has not been assessed in detail. 
    
The fragmentation function $D_{i}$ for each hadron species is necessary for our template calculation, hence any uncertainty on the $D_{i}$ induces an uncertainty on the extracted $m_{t}$. There are many possible reasons for uncertainty on $D_{i}$, starting from the experimental data that are used to measure it, the theoretical quantities and methods involved in the extraction of $D_{i}$ from the data, and more practical details such as the possible sensitivity to kinematics and color environment between the experiments in which the measurement is carried out and those in which the function is used. (For example, the use of $e^{+}e^{-}$ data to extract $D_{i}$ to be used at a hadron collider.) 
 
It is not possible for us to list all possible sources of uncertainty on $D_{i}$, which depend crucially on the method used to obtain such functions. In order to obtain a meaningful and useful result we aim at characterizing the shape of $D_i$ using  the first few Mellin moments of the $D_{i}$ function. This kind of characterization is analogous to the characterization of probability density functions through their mean, standard deviation, skewness and kurtosis. Indeed, we define the Mellin moments of $D(z)$ such as $\langle z \rangle$ is the first moment, and  $\int z^{i}D(z)\,dz$ is the $i$-th moment.\footnote{We recall that, in principle, from the infinite series of Mellin moments, one would be able to faithfully describe the $D_i$ function in Mellin space, but we limit the calculation to the first few coefficients, as the extraction of high Mellin moments is usually challenging from both experimental and theoretical points of view.}

Armed with this characterization of $D_i$ in terms of Mellin moments, we compute the sensitivity of the extracted $m_{t}$ to the first three the Mellin moments. In this way any possible change in $D_{i}$ due to future, improved theoretical tools used in their extraction, or better data, or any correction factor which may be used to account for kinematic or color effects can be cast as a shift of the Mellin moments and translated into a shift in the top quark mass.

\begin{figure}
        \centering
        \includegraphics[width=0.59\textwidth, trim = 0 0 0 30, clip]{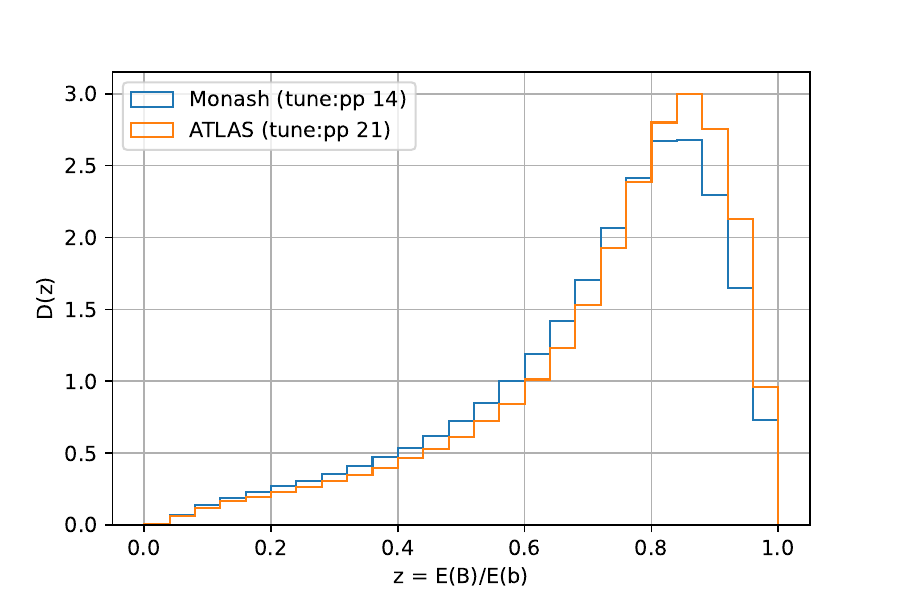}
        \caption{Fragmentation function obtained from Pythia~8 using a ATLAS tune (Tune:pp 21) and the default tune (Tune:pp 14) for   $m_t=171$~GeV.}
        \label{fig:frag_func_tunes}
    \end{figure}

To probe the dependence of our result on the Mellin moments of the fragmentation functions $D_{i}$, we perform two analyses. In each analysis, we compare two or more physical setups, each of which is internally consistent and physically distinguishable from the others and which, in particular, yields a different fragmentation function. In the first analysis, we compare two different tunes of Pythia~8.2, namely the default tune of Monash (Tune:pp 14) and a tune by ATLAS derived from Monash (Tune:pp 21). These two tunes give different $D_{i}$ functions shown in Figure~\ref{fig:frag_func_tunes}. The difference in the  Mellin moments of these two tunes is given in  Table~\ref{tab:tune_frag_sensitivity} and results in a change in the extracted top quark mass well above a few GeV. Such a large difference is the result of the current poor knowledge of the dynamics of hadronization when compared with the ambitious goals of the top quark mass extraction. In order to do a more tailored job in our assessment of the sensitivity to $D_{i}$, we produce a new $D_{i}$ whose value at each $z$ is given by the Monash tune weighted at 90\% and the ATLAS tune weighted at 10\%. The weighted average of $D_{i}$ functions that we obtain from the two tunes results in a  difference in the extracted top quark mass of 1.7~GeV, thus it is better suited to mimic a situation where two future determinations of $D_{i}$, improved over the  current ones, will be compared in a top quark mass extraction. The resulting change in the extracted $m_{t}$ is used to compute a sensitivity ratio as in Eq.~(\ref{eq:sensitivity-ratio}), from which we observe that a knowledge of the (Mellin moments of) $D_{i}$ with precision about 3 times better than the desired precision on $m_{t}$ is needed if one does not want to see the $m_{t}$ measurement spoiled by poor knowledge of $D_{i}$.

\begin{figure}
        \centering
        \includegraphics[scale=0.6, trim = 0 0 0 30, clip]{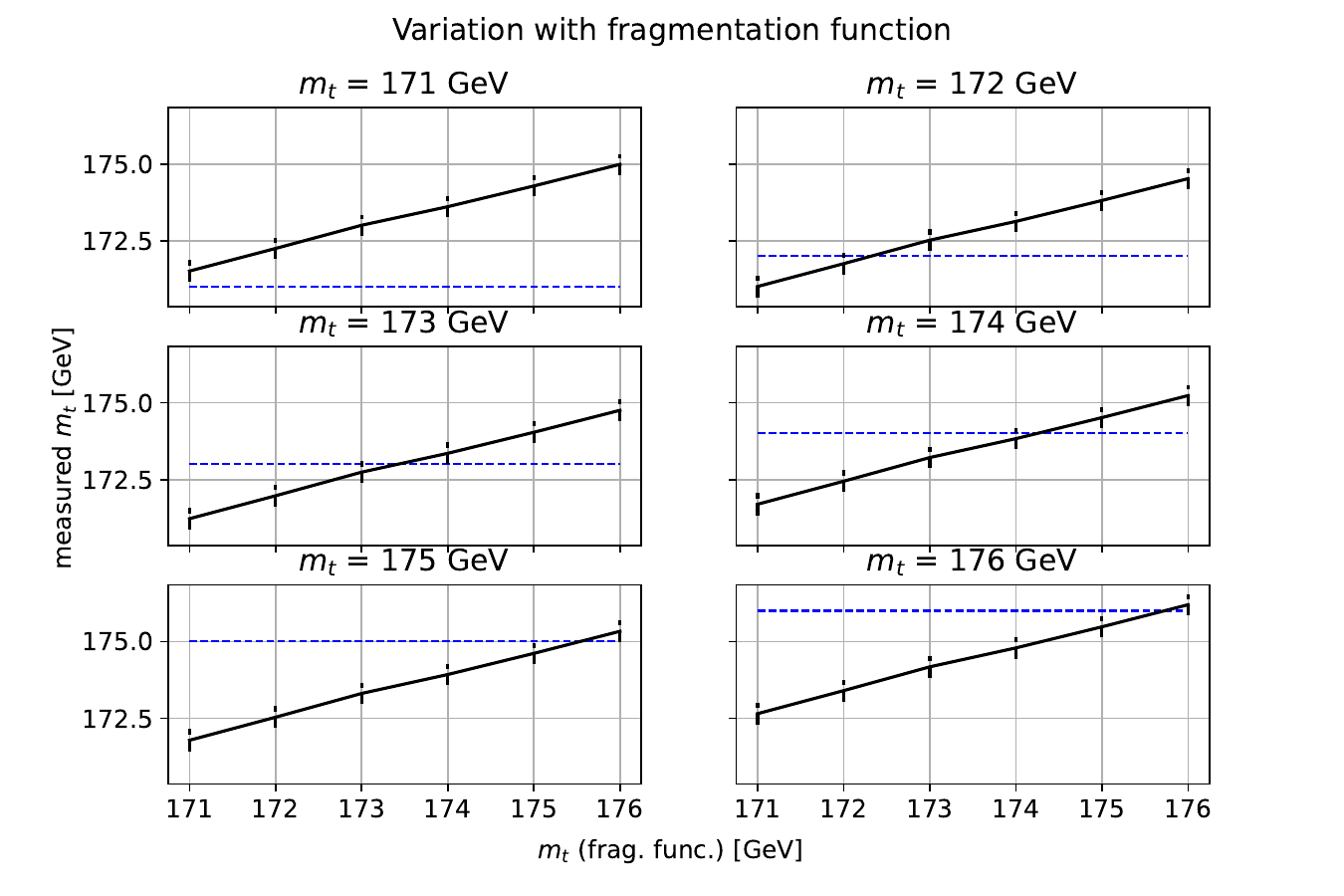}
        \caption{Effect on  the extracted $m_{t}$ from the change of the fragmentation function as parameterized by changing $m_{t}$ in the data used to the MC truth on which the fragmentation functions is measured. The $m_t$ used to measure the fragmentation is on the horizontal axis;  the measurement is shown as a black line for each subplot  corresponding to a correct $m_t$ used to generate data. The blue line is shown as a reference, as it  corresponds to an unbiased measurement. }
        \label{fig:var_frag_func}
    \end{figure}

This conclusion is corroborated by the second analysis in which we exploited the fact that in our procedure we have extracted $D_{i}$ function directly from Monte Carlo truth, namely, we have extracted a set of $D_{i}$ functions for each hadron species and for each $m_{t}$ for which we have generated Monte Carlo data. Therefore we can mimic a mismatch between the $D_{i}$ functions appropriate for the data and a slightly incorrect one used in the computation of templates by simply trying the extraction of $m_{t}$ from one data sample using templates Eq.~(\ref{ansatz_LB}) based on $D_{i}$ that correspond to a different value of $m_{t}$. The result of several exercises of this type is reported in Figure~\ref{fig:var_frag_func} where we show the extracted $m_{t}$ for several input values and several choices of the $D_{i}$. The choices of the $D_{i}$ are labeled by $m_{t}$ values along the horizontal axes. In this case, we choose to parameterize the effect by looking at the Mellin moments of the $D_{i}$ functions. 
 
We summarize these findings and quote our result as a sensitivity reported in Table~\ref{tab:frag_sensitivity_mt}. As for the first analysis, we find that  a knowledge of the (Mellin moments of) $D_{i}$ with precision about 3 times better than the desired precision on $m_{t}$ is needed if one does not want to see the $m_{t}$ measurement spoiled by a poor knowledge of $D_{i}$.

Other analyses are possible to asses the sensitivity to $D_{i}$.  For instance, one could try to use $D_{i}$ computed from experimental inputs using calculations to some order in perturbation theory, e.g. following recent efforts to improve the knowledge of fragmentation from first principles calculations \cite{ Fickinger:2016rfd,Maltoni:2022bpy,Czakon:2021ohs,Czakon:2022pyz} which contain scale parameters, such as the fragmentation and renormalization scales. As customary in pQCD studies the change on $D_{i}$ that follows from the variation of these scales can be taken as a source of uncertainty in the results, in our case on the templates Eq.~(\ref{ansatz_LB}) and the $m_{t}$ extraction to which they give rise. We do not attempt this kind of analysis here, as it has more to do with the work of extraction of $D_{i}$. Furthermore, effects on $D_{i}$ from this type of theory uncertainty can be captured in any case by our statement on the Mellin moments described above. 

It is important to remark that the present knowledge of $D_{i}$, be it taken as a phenomenological parameter tuned over data, e.g one of the Pythia tunes we have compared, or from first principle computations applied on data, is far from the necessary precision to not spoil a determination of $m_{t}$ that aims at a sub-GeV precision. As far as we know this situation is common to other methods that explicitly use hadrons produced in $t\bar{t}$ events, see e.g. \cite{Corcella:2017rpt,Agashe:2016bok,CMSlifetime,ATLAS:2022jbw,ATLAS:2019ezb}. Therefore we urge for improvements on the determination of fragmentation functions both on the experimental and theoretical side.
    
    \begin{table}
    \begin{centering}
    \begin{tabular}{|c|c|c|c|}
    \hline 
    \hline
    Mellin Moment & $\delta \langle z^n \rangle / \langle z^n\rangle $ & $\delta m_t (10\% \text{ reweighting} ) $ & Sensitivity \tabularnewline
    \hline
    $\langle z\rangle $ & 2.8 \% & & 3.5\\
    \cline{1-2} \cline{4-4}
    $\langle z^2\rangle $ & 5.2 \% & ~ 1.7 GeV & 2.5\\
    \cline{1-2} \cline{4-4}
    $\langle z^3\rangle $ & 7.2 \% &   & 1.4\\
   \hline
   \hline
    \end{tabular}
    \par\end{centering}
 
    \caption{\label{tab:tune_frag_sensitivity} For each of the first three Mellin moments of the  $D_{i}$ we report: the difference between the default Pythia tune (Tune:pp 14) and the ATLAS tune (Tune:pp 21); the effect on the extracted $m_{t}$ stemming from a 10\% contamination of the ATLAS tune into the Monash tune; the sensitivity of the extracted $m_{t}$ to each Mellin moment. }

    \end{table}
    \begin{table}
    \begin{centering}
    \begin{tabular}{|c|c|c|c|}
    \hline 
    \hline
    Mellin Moment & $\delta \langle z^n \rangle/\langle z^n \rangle$ & $\delta m_t^{{(171\to 176)}} $ & Sensitivity \tabularnewline
    \hline
    $\langle z \rangle$ & 0.53 \% & & 3.8\\
    \cline{1-2} \cline{4-4}
    $\langle z^2 \rangle$ & 0.91 \% & ~ 3.5 GeV & 2.2\\
    \cline{1-2} \cline{4-4}
    $\langle z^3 \rangle$ & 1.23 \% &   & 1.6\\
   \hline
   \hline
    \end{tabular}
    \par\end{centering}
    \caption{\label{tab:frag_sensitivity_mt} For each of the first three Mellin Moments of the fragmentation function we report: their change due to varying  the $m_t$ value that labels the $D_{i}$ extracted from the Monte Carlo truth from 171 GeV to 176 GeV;  the change on the extracted $m_{t}$ due to using the $D_{i}$ extracted from the Monte Carlo truth for $m_{t}=176$~GeV on the data sample for $m_{t}=171$~GeV; the sensitivity of the extracted $m_{t}$ to each Mellin moment.}
    \end{table}

\subsection{Uncertainties due to top quark production modeling} 

As mentioned earlier, CDF and CMS have already implemented a method to measure top quark mass using $B$-hadron decay length~\cite{Abulencia:2006rz,CMSlifetime}. More specifically, they use the mean transverse decay length of $B$-hadrons in top quark events. Like our method, their results are not plagued by jet energy scale uncertainty. The largest uncertainty in the latest CMS analysis is due to modeling of the top quark $p_T$ distribution~\cite{CMSlifetime}, which shows how the measurement is in general sensitive to the production mechanism of the top quarks. Our method, if applied on fully inclusive data samples and with a perfect ansatz for the shape of the $b$ quark energy distribution, would be independent of top quark production mechanism, hence we expect our results to have little sensitivity to top quark $p_T$. In fact, some degree of sensitivity is unavoidable because we use only part of the data and our ansatz is unavoidably not perfect. Indeed we tune hyper-parameters by doing numerical experiments to minimize the bias of the method and we do that assuming Standard Model production. This procedure and the event selection cuts could lead to a non-zero sensitivity to top quark $p_T$ and to other aspects of the production mechanism of the top quark in general. In the following, we describe how we probed this effect and we concluded that our method is indeed practically insensitive to top quark $p_T$. 

In order to quantify the sensitivity of our method to changes in the top quark $p_{T}$ distribution we perform the following exercise, which also gives us the opportunity to compare our method with the $L_{xy}$ methods of CMS in this particular respect.
To carry out extraction of $m_{t}$ from the $L_{xy}$ spectrum we generate the $L_{xy}$ distribution with the same event selection criteria as we use earlier for our method. In order to learn the response of the $L_{xy}$ distribution to changes of $m_{t}$ we generate event samples for 7 different values of $m_t$ from 170~GeV to 176~GeV and we fit each bin count of the {$L_{xy}$} distribution using a quadratic polynomial as a function of $m_t$.  Having learned how each bin count of the $L_{xy}$ distribution depends on $m_{t}$ we generate a sample for a test value of $m_{t}$, that we call $m_{t,test}$ and we obtain a top quark mass measurement from a $\chi^2$ minimization.
As a check of the goodness of this procedure, we checked that this method returns $m_{t,~true}$ free of any bias. 

\begin{figure}
\centering
\includegraphics[scale=0.6]{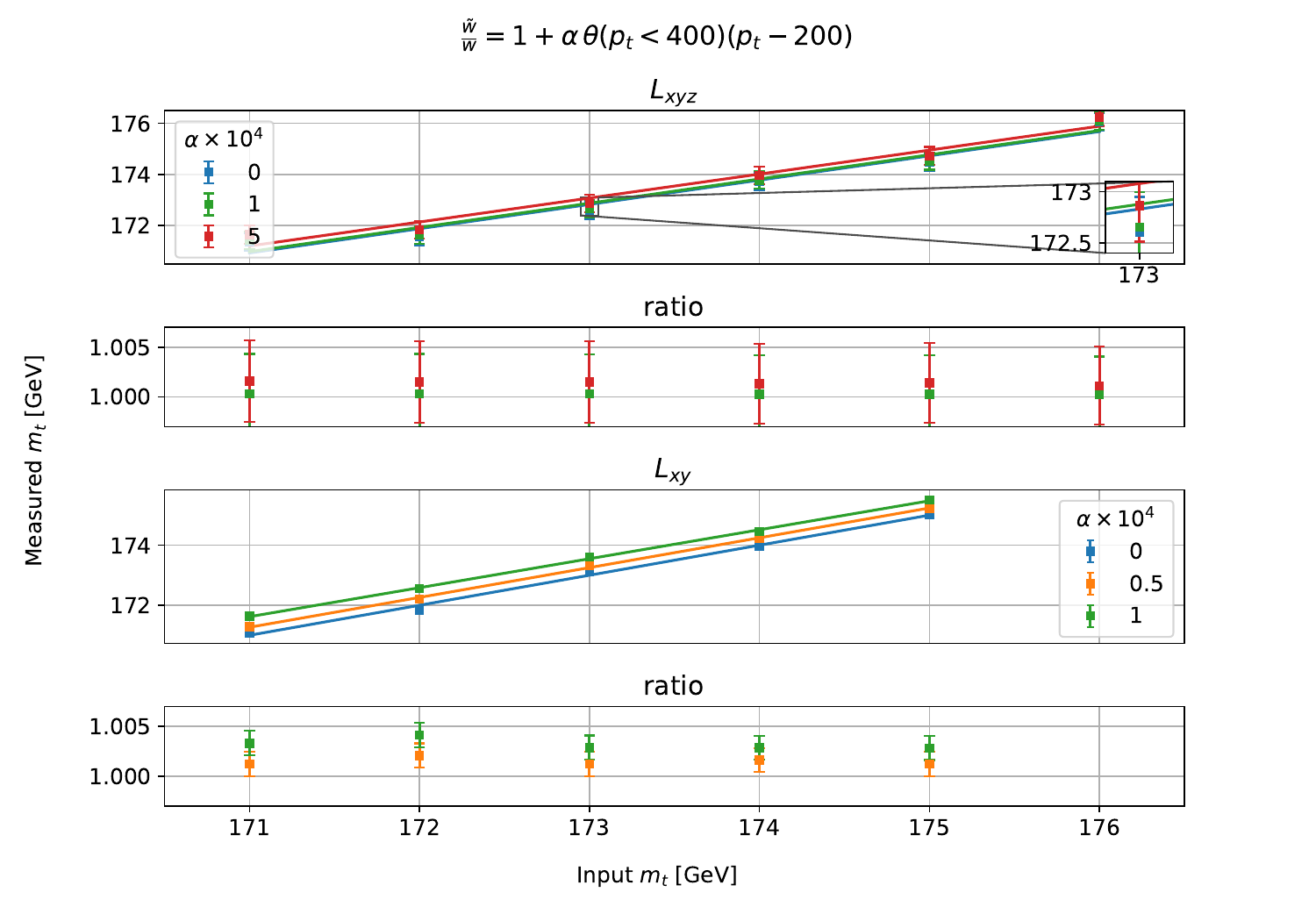}
\caption{Bottom two plots are the result of $L_{xy}$ method when re-weighting the top quark $p_T$ distribution. The ratio plot shows the ratio of the predictions before and after re-weighting. The green line/data points represent a $\sim 0.5 \%$ change in the average top quark $p_T$. The top two plots show the results of our method with an inset included to show how small the shift is for the same re-weighting of top quark $p_T$ distribution. It should be noted that the ratio due to re-weighting for $\alpha = 10^{-4}$ is on average smaller than 1.001 for our method, compared to $\sim 1.004$ for the $L_{xy}$ method.}
\label{fig:moneyplot}
\end{figure}

For a fixed luminosity we find that the statistical error obtained on the $m_{t}$ measurement from the $L_{xy}$ spectrum is better than the result of our method in Eq.~(\ref{eq:money}). However, the transverse decay length method is highly sensitive to top-quark $p_T$ compared to our method. We quantify the sensitivity of the transverse length method and of our method by performing the top mass extraction on $L_{xy}$ and $L_{xyz}$ spectra, respectively, obtained by reweighting each event of the $m_{t,test}$ sample according to the top quark  $p_{T}$ in that event. The new weights are obtained using 
\begin{equation}
    \text{w}_{\text{new}} = \text{w}_{\text{old}} \cdot \left[ 1 + \alpha \cdot \Theta(p_T < 400)\cdot (p_T -200) \right],
\end{equation}
where $\text{w}_{\text{new}}$ and $\text{w}_{\text{old}}$ are the new and the old weights, $\Theta(c)=1$ when $c$ is true and $\Theta(c)=0$ otherwise, and $\alpha$ adjusts the amount of re-weighting. We stop re-weighting at a $p_{T}$ threshold of 400 GeV, as there are fewer data at such large $p_{T}$, hence the impact on the measurement is modest, and the entailed larger statistical uncertainty makes it less meaningful to manipulate the distribution at such large $p_{T}$.
Figure~\ref{fig:moneyplot} shows how the two methods compare. It is clear that even for a very soft re-weighting, the transverse decay length prediction shifts appreciably, whereas predictions using the full decay length fitted over our templates Eq.~(\ref{ansatz_LB}) do not move nearly as much. 
For concreteness one can look at the re-weighting for $\alpha=10^{-4}$, which roughly corresponds to the theoretical uncertainty in top $p_T$ spectrum shown in Ref.~\cite{CMS:2021vhb}, this is roughly moving the average top quark $p_T$ by around 0.5\%.  At this value of $\alpha$, the prediction using $L_{xy}$ method shifts by $\approx 600$~MeV, whereas for $L_{xyz}$ the shift is only $\approx 50$~MeV.

\section{Conclusions}
\label{conclude}

The LHC and its High-Lumi upgrade face the formidable task of improving our understanding of the Standard Model, which is already very well established. Nevertheless, there are corners of the theory that are in urgent need of further study. The existence of multiple proposals for a Higgs ``factory'', or an even more versatile factory that can also produce top quarks and electroweak bosons is a clear indication of the need to sharpen our knowledge of the Standard Model with higher precision and potentially see discrepancies resulting from new physics.

The top quark mass is a very special parameter in this respect because it involves a number of perturbative and non-perturbative issues on the theory side, as well as a variety of experimental challenges resulting from the relatively complex decay modes of top quarks. At present, there are measurements that claim sub-GeV precision for the top quark mass, but further scrutiny and caution are needed because of the myriad delicate physical phenomena that affect how the process is manifested in the data, how the complex data are analyzed by the experimental teams, and finally how this is all translated into a theoretical interpretation. In view of this, and the long period ahead of us before a lepton collider will be ready to study top quarks, there is an urgent need to develop new methods to determine the top quark mass at the LHC. These methods can also be applied to advantage at future colliders. The hope is that, by getting a number of high-precision measurements, with different machines, experiments, and methods with  different and complementary sources of systematic uncertainty, a coherent picture will emerge that provides greater clarity and a detailed understanding of the delicate phenomena alluded to above in order to have a similarly clear and robust measurement of $m_{t}$.  For this program to take place, it is necessary to diversify the types of observables and the theory approaches used to extract $m_{t}$. 

With this in mind, we proposed a new observable that is essentially the transposition of the energy peak proposal of Ref.~\cite{Agashe:2012bn} from the energy domain into the decay length domain. The shift from an energy observable to a measured length has as most dramatic effect the elimination of much of the jet energy scale uncertainty from the list of experimental   uncertainties for the measurement. 

The flip side  of this proposal is the increase of the importance played by theoretical uncertainties tied to the description of the formation of long-distance objects, i.e. the mesons and hadrons whose decays we observe in detectors.
This phenomenon tracks back to the great mystery of confinement dynamics in QCD. The difficulty and depth of this issue have made it such that it has been only possible to present phenomenological descriptions for the transition of quarks into hadrons, even after many decades of perturbative QCD being the established theory of strong interactions. While a deeper understanding of hadronization may be attained in the future, we have investigated how much the present ignorance of the dynamics of hadron formation would impact the extraction of the top quark mass using a method based on hadron decay length measurements.

To carry out our study we used the Monte Carlo event generator Pythia~8.2 and we explored a number of aspects of the description of the parton shower and hadronization models of this type of event generator. The convenience of using an event generator allowed us to mimic the same process for the extraction of $m_t$ that the experiments would carry out on real data. Following a standard procedure, we extract $m_t$ by minimizing a $\chi^2$ obtained by comparison of  template length spectra that we compute with Pythia~8.2 for several values of $m_t$ and separate spectrum, for a simulated sample that is also based Pythia~8.2, which plays the role of the real data. 

The observable that we use to extract $m_t$ is the three-dimensional $B$-hadron decay length, $L_{xyz}$, and we  carry out the $\chi^2$ analysis using a carefully crafted template that is given in Eq.~(\ref{ansatz_LB}). This template returns the distribution of the $L_{xyz}$ observables for a given $m_t$, exploiting the fact that the average distance traveled by a $B$-hadron in the laboratory has a one-to-one correspondence with the energy of the $b$-quark that originated the $b$-flavored hadron. The key for this link to exist is that the distribution of the energy of the $b$-quark is, in turn, related to $m_t$ by the invariance of the peak position discussed in Ref.~\cite{Agashe:2012bn}. With this chain of links between observables in mind, we have formulated Eq.~(\ref{ansatz_LB}) as a candidate template for the measurement of $m_t$. We have verified that a measurement of $m_t$ can be carried out with this template without introducing a significant bias. We have also verified that a number of ``hyper-parameters'' of 
 our template can be set to the values collected in Table~\ref{tab:hyper-parameters-best} that are optimal for the stability of our result, maintain negligible bias, and insure a competitive uncertainty in the measurement of $m_t$ when $\mathcal{O}(100/{\rm fb})$ or more of LHC data are used in the analysis.

Having established that a robust and precise measurement of $m_t$ can be carried out with templates of the form Eq.~(\ref{ansatz_LB}) on the $L_{xyz}$ measurements, we then stress-tested our method with the aim of assessing the sensitivity of our result to the many theoretical and experimental quantities that are used to compute our templates.
 We have tested in particular how well the dynamics of parton shower and hadronization that determines the fragmentation functions $D_i$ needs to be known in order not to spoil the precision of this method. With two different assessments, summarized in Tables~\ref{tab:tune_frag_sensitivity} and \ref{tab:frag_sensitivity_mt}, we showed that fragmentation functions known at the sub-percent level are needed. Indeed fragmentation dynamics has emerged as the most important challenge to attain an uncertainty in $m_t$ at the 500~MeV level.
As a potential second challenging source of uncertainty, we have identified the hadronization fractions, which account for the abundance of each hadron species produced in the phase-space covered by the experiment. The expected theoretical uncertainty from the hadronization fractions, the fragmentation function, and collected statistics at the LHC are given in  Eq.~(\ref{eq:money}), which formulates our final predictions for the uncertainty on $m_{t}$ in terms of the levels of uncertainty in the hadronization functions and fractions.

It is no accident that we have not included in our summary of the most prominent uncertainties a possible source of uncertainty from the mis-modeling of top quark production. In fact, we tested explicitly how much our extracted top quark mass would change if the transverse boost of the top quark in the data were to differ from that predicted by the event generator prediction and we  found that even for a mis-modeling larger than the current NNLO theory error band, the uncertainty on $m_t$ would still be negligible compared to the LHC target of few hundred MeV. This is not at all true of the method that comes closest to our proposal, which is the CMS measurement \cite{CMSlifetime} that makes use of $L_{xy}$, the decay length projected onto the plane transverse to the beam axis. As in our method, the choice of decay length is motivated by the elimination of jet energy scale uncertainties but in the CMS case, templates based on any invariance of the underlying $b$-quark energy are not employed. As a consequence, the CMS result is very sensitive to the top quark production kinematics. Indeed, the uncertainty related to top quark transverse boost was found to be the largest theoretical uncertainty, exceeding 1~GeV in Ref.~\cite{CMSlifetime}. The striking reduction  of sensitivity to top quark transverse momentum in our method is seen in Figure~\ref{fig:moneyplot}, highlighting the benefit of decay length distribution templates rooted in energy-peak invariance.

We stress that our method may be inferior to other proposals when one looks solely at the statistical part of the uncertainty. Indeed we found that our implementation of the CMS method based on $L_{xy}$ gives a smaller statistical uncertainty than our method.
However, given the extremely large sample of top quarks that has already been recorded by the LHC experiments and the even larger sample that will be collected at the High-Lumi LHC,  statistical uncertainty will not be an issue and essentially all of the LHC measurements will ultimately be limited by systematic uncertainties. 

Further study is necessary to transform our proposal into a real measurement of $m_{t}$. For sure, a thorough study of the measurement uncertainties for the lengths would be necessary. Indeed each $B$-hadron species is detected in different decay modes, whose detection may be more or less amenable to the measurement of the three-dimensional decay length. On a related topic, one can imagine that each species of $B$-hadrons being measured with different accuracy, the measured length spectra for each hadron species can be translated into a ``particle-level'' length spectra, in which the effects of detector mismeasurement have been removed.  This procedure is applied routinely in SM studies at the LHC, especially for kinematic properties of SM reactions, and goes by the name of detector unfolding. Given the different role played by each hadron species and potentially different degrees of understanding of the fragmentation function $D_{i}$ and the hadronization fractions $f_{i}$ for each species, it might be useful to explore how to carry out this type of measurement on unfolded data for each hadron species.

Another branch of development that is worth mentioning is a possible reduction of complexity in the computation of templates Eq.~(\ref{ansatz_LB}). This type of ansatz requires a double integral, one of which is over $E_{B}$, and in part serves the purpose of convolving the exponential decay law with the result of the integral over $E_{b}$. As the exponential decay law is a piece of extremely well-established physics, one could potentially consider it a part of the detector smearing so that the measurement of a length can be transformed into a measurement of average decay length, or, if one wishes, a measurement of $E_{B}$, through the relation $\lambda_{B_{i}}=\langle L_{i} \rangle = \gamma \beta \cdot c \tau_{i,0} $ where $\gamma=E_{B_{i}}/m_{B_{i}}$ and $\gamma=1/\sqrt{1-\beta^{2}}$. 

While it was not an issue for our work to generate adequately many sets of precise templates from the double integral Eq.~(\ref{ansatz_LB}) it might become a more important issue in an actual measurement with collider data. As a matter of fact, the $E_{B}$ integral is rendered more complicated by the presence of the exponential, which plays the detrimental role of ``diluting'' the information and forces us to perform a more complicated integral to undo its effect for each and every template that we compute. Therefore, one can envisage being more economical and undoing the exponential part of the integral once and for all in the unfolding. Such an observation may enable a quicker path to more thorough explorations of showering and hadronization effects or changes in top quark production kinematics for which it may be necessary to generate numerous dedicated sets of templates. Amusingly, the unfolding of the exponential convolution from $E_{B}$ to $L_{B}$ space could be attempted using semi-analytical techniques for the inversion of Laplace transforms~\cite{Stehfest_1970,CohenLaplace}.

\section*{Acknowledgements}

K.~A.~and J.~I.~thank O.~W.~(Wally) Greenberg  for organizing the ``50 Years of Quarks and Color" symposium at the University of Maryland, where K.~A.~and J.~I.~first discussed the basic idea in this paper.
The authors acknowledge Snowmass 2021, as it motivated them to make significant progress in this work. 
J.~I.~acknowledges the support of DOE grant DE-SC0011702 and the Joe and Pat Yzurdiaga endowed chair in experimental science.
The work of K.~A., S.~A. and D.~S.~is supported by NSF Grant No.~PHY-2210361 and by the Maryland Center for Fundamental Physics. 
The work of D.~K. is supported by the DOE Grant No. DE-SC0010813.




\end{document}